\documentstyle[12pt,aas2pp4]{article}

\received{}
\accepted{}
\journalid{}{}
\articleid{}{}
\lefthead{Robertson S.L.}
\righthead{The Intrinsic Magnetic Fields of the Galactic Black Hole Candidates}

\begin{document}
\title{The Intrinsic Magnetic Fields of the Galactic Black Hole Candidates}
 
\author{STANLEY L. ROBERTSON}

\affil{Department of Physics, Southwestern Oklahoma State University,
Weatherford, OK 73096}

\begin{abstract}
Recent work has linked the quiescent luminosities and hard/soft spectral
state switches of neutron stars (NSs) to their spinning magnetic fields.
It is shown here that the quiescent luminosities and
spectral state switches of galactic black hole candidates (BHCs) could be
produced in the same way for spin rates below 100 Hz and magnetic fields above
$10^{10}$ G. It is also shown that the ultrasoft peaks and large flickering
amplitudes of the BHCs would be expected from the surfaces of massive NSs.
None of the few spectral characteristics that distinguish BHCs from low mass
NSs have been explained in terms of event horizons.
Serious consideration of the possibility that they might simply be
massive NSs opens an avenue for proof of event horizons by negation, but
requires the use of a space-time metric that has no event horizon. The Yilmaz
exponential metric used here is shown to have an innermost marginally stable
orbit with radius, binding energy and Keplerian frequency that are within a few
percent of the same quantities for the Schwarzschild metric. A maximum NS mass
of $\sim 10 M_\odot$ is found for the Yilmaz metric. The two metrics
essentially differ only by the presence/absence of a surface for the BHCs,
thus enabling proof or disproof of the existence of event horizons.
\end{abstract}


\section{Introduction}
According to General Relativity,
an object of nuclear density and more than 2.8 M$_\odot$ would
be a black hole (Kalogera \& Baym 1996, Friedman \&
Ipser 1987). Several black hole candidates (BHCs)
have been found in x-ray binary systems.
At this time there are ten galactic x-ray sources known to exceed the
neutron star (NS) mass limit. Only two of these, Cyg X-1 and LMC X-3, are not
in low mass binary systems. Perhaps twenty more candidates have been identified
via spectral similarities (e.g. see Barret, McClintock \& Grindlay 1996).
Active galactic nuclei (AGNs) are also generally believed to
exceed a Schwarzschild mass limit, however, they will be
considered here only in passing in order to restrict the scope of a lengthy
article. In addition, their radii are less well constrained than those of the
x-ray novae. Nevertheless, it should be noted that there are strong spectral
similarities between NSs, BHCs and AGNs. Campana et al. (1998a), Becker
\& Tr\"{u}mper (1998) and van der Klis (1994) have recently reviewed
NS properties, while Poutanen (1998) has reviewed BHCs and AGNs.

X-ray novae display a rich mix of spectral and timing characteristics.
Many of them are keyed to the luminosity level. Luminosity
generally derives from accretion of mass from a companion star
via an accretion disk.
Flares are caused by instabilities of the flow either from
the companion or through the disk. Maximum luminosities during major
flares are often $\sim 10^6 - 10^7$ times quiescent
luminosities. Maximum luminosities are often near the
Eddington limit ($\sim 2 x 10^{38}$ erg/s for a canonical
1.4 M$_\odot$ NS). Complex changes of
luminosities in soft x-rays (0.1-4.0 keV), hard x-rays (4-20 keV) and hard
tails, (20-200 keV) occur during flares. Hard tail
luminosity with an inverse power law dependence on
photon energy is prominent during the initial rise of luminosity and also
during the decay. In most cases this power law component weakens substantially
at high overall luminosities. ``Ultrasoft'' radiation (White \& Marshall 1984)
with a general brehmsstrahlung shape and a peak near 1 - 3 kev
is often, but not always, seen at high luminosities. Observations during
``dips'' where some sources, seen at large inclination through
absorbing materials, provide some spatial resolution, show that the
soft peaks arise near the center of the the disk.
BHCs simultaneously show hard tails and softer, more luminous
peaks more commonly than NSs. Spectral state switches occur between
soft ``high'' luminosity states and the harder ``low'' states that are
characterized by power law spectral features. Hard low states typically
have luminosities below $10^{36}$ erg/s for NSs and below $10^{37}$
erg/s for BHCs. Campana et al. (1998) and
Zhang, Yu \& Zhang (1998) have attributed the spectral state switch to
``propeller effects'' (Illarianov \& Sunyaev 1975) of magnetic
fields for NSs. Black hole models usually attribute the switch to
accretion disk instabilities. In either case, it is generally
accepted that the hard spectrum of the low state originates
in the accretion disk and that it persists, usually becoming harder,
as quiescence is approached. There is no consensus
about how it is produced. The strong similarities of quiescent
NSs and BHCs (Tanaka \& Shibazaki 1996) suggest that a common mechanism
produces the quiescent luminosity.

Low frequency, $\approx$ 6
Hz, quasi-periodic oscillations (QPOs) often appear in
suitably high states (Fortner, Lamb \& Miller 1989, Miyamoto,
Kimura \& Kitamoto 1991, Makishima et al. 1986) of both NSs and BHCs.
Large amplitude flickering ($\approx$ 10$^{-3}$ - 10 s luminosity
variation) occurs when power law components are dominant, particularly at
intermediate luminosity (Balucinska-Church et al. 1997, Tennant, Fabian \&
Shafer 1986, Stella et al. 1985).  The variations of
hard and soft x-rays are often correlated (usually with delayed
hard photons) on short time scales ($< 1 $ s), but are
uncorrelated (Pan et al. 1995) or anti-correlated on scales of hours or more.
All of these features, some of which were once thought to
be black hole signatures, have been observed in both
BHCs and NSs (e.g., see Singh et al. 1994, Barret, et
al. 1992, Churazov, et al. 1995, Stella et al. 1985, Tanaka 1989,
van der Klis 1994). Cir X-1, which was dropped from the BHC list after
displaying surface thermonuclear bursts, shows most of these
characteristics. Its wide array of spectral characteristics is
likely due to a rather wide variation
of accretion rates associated with an eccentric orbit.

Some NSs show some features that are not obviously shared
by BHCs. Radio or x-ray pulsations, and surface thermonuclear
bursts are NS signatures that have only been found in
erstwhile BHCs such as Cir X-1. NSs usually
show bursts or pulses but not both. The accepted explanation for
this dichotomy is a magnetic field, a NS signature, that can concentrate
accreting matter in magnetic pole regions where fusion proceeds
at rates high enough to suppress bursts (Taam \& Picklum 1978).
Bursts are generally not observed for pulsars with $\sim 10^{11}$ G
magnetic fields, but have been observed for millisecond pulsars,
which have weaker fields.
Interaction of the accretion flow with the NS magnetic field is believed to
account for the characteristics of color-color diagrams  and
hardness-intensity diagrams of ``Atoll'' and ``Z'' type NSs.
These sources have magnetic fields of $\sim 10^8$ G and $10^9$ G, respectively
(White and Zhang 1997). The peak accretion rates of Zs are near Eddington
limits, while those of Atolls are much lower.
Van der Klis (1994), who developed much of the Atoll-Z phenomenology has
provided a review and a discussion of the similarities between
the various states of BHCs and NSs; particularly the strikingly similar
time-resolved power density spectra of flickering Atolls and BHCs.

The presence of a hard spectral tail at overall luminosity levels
above $10^{37}$ erg/s is considered to be a reliable signature
of a BHC (Barret, McClintock \& Grindlay 1996). Why an accreting
black hole should have this property is not understood
(Kusunose, Minishige \& Yamada 1996, S.N. Zhang, et al. 1997).
With this exception, the lack of spectral or timing signatures of
black holes has left mass determination as the only way
to certify a BHC (McClintock 1998). Until signatures of event horizons
are found, it remains feasible to suppose that the BHCs might simply be
massive neutron stars. The point of this work is to explore this possibility.

BHCs as compact, massive NSs could possess magnetic fields, and should have
large surface binding energies and substantial redshifts of surface radiations.
Magnetic fields would permit a common mechanism for the spectral state
switches of BHCs and NSs. Larger surface binding energies and redshifts
for the more massive BHCs would make their surface emissions both
brighter and softer. It would seem necessary to reject these possibilities
before acceptance of the reality of event horizons for the BHCs.
In order to pursue these possibilities in a quantitative way,
a space-time metric that has no event horizon is needed.
The Yilmaz metric (YM) has been adopted here for this purpose. The massive NS
hypothesis discussed here is only weakly dependent on metric properties
in general, but depends crucially on the distinctions between surfaces
and magnetic fields \textit{vs} event horizons. It is shown here that YM and SM
accretion disks have such similar properties that they essentially
differ only by the presence/absence of a surface.

Although most of the mathematical details have been placed in an appendix, a
comparison of the two metrics is given in Section 2. Results for
the two metrics for a simple model of a star of constant proper density
(Clapp 1973) are also given. An object of this sort with a density in its outer
layers corresponding to nuclear saturation density provides a rough
approximation of a neutron star with a fairly stiff equation of state. It
provides very conservative estimates of the radiant energy and redshifts
to be expected for accretion reaching a massive NS surface.
Magnetic field phenomena that affect the accretion flow and spectral
characteristic are outlined in Section 3.
Section 4 discusses magnetic and strong-field gravitational effects in
NSs and BHCs. Section 5 considers the origins of various
spectral characteristics and the implications for massive
NS models.

\section{The Yilmaz Metric}
While not widely known, the Yilmaz theory (Yilmaz 1958, 1971, 1975, 1992,
1994, 1995) has some interesting positive features that may help to
resolve the hiatus between General Relativity and Quantum Mechanics.
At this time, it is quite clear that one or both of these landmark theories
of the twentieth century must be modified before they can be fully
compatible. The Yilmaz theory modifies General Relativity primarily by the
explicit inclusion of the stress-energy of the
gravitational field as a source of space-time
curvature. With a true field stress-energy the Yilmaz
theory possesses a field Lagrangian and can be quantized (Yilmaz 1995).

Although the Yilmaz theory has been criticized, (Will 1981) the criticism
appears to be based on an incorrect assumed form of the metric (Yilmaz 1981).
For the present purposes, the essential features of any alternative metric
would be lack of event horizons and readily calculable NS characteristics.
Nevertheless, the Yilmaz theory passes the four classic
weak-field tests. It permits local energy-momentum conservation and
has no adjustable parameters, no singularities and no
event horizons (Alley 1995, Yilmaz 1994). Gravitationally compact
objects can exist in the YM but they are not black holes. Radially directed
photons can always escape.

The metric of space-time in the vicinity of
neutron stars is dominated by the gravity of the star which is considered
here to be a static object. The static limit interval in the
Yilmaz exponential metric is:
\begin{equation}
ds^{2}~=~g(r)c^{2}dt^{2}~-~(dr^{2}~+r^{2}d\theta^{2}+r^{2}\sin
^{2}\theta d\phi^{2})/g(r)
\end{equation}\\
where:
\begin{equation}
g(r)~=~\exp(-2 u(r))
\end{equation} and u(r) is the gravitational potential. In the SM:
\begin{equation}
ds^{2}~=~g(r)c^{2}dt^{2}~-~dr^{2}/g(r)~-~r^{2}d\theta^{2}~-~r
^{2}\sin^{2}\theta d\phi^{2}
\end{equation} and:
\begin{equation}
g(r)~=~1~-~2 u(r)
\end{equation}
The gravitational potential at distance r from mass M, $u(r)=GM/c^2r$
reaches $\sim 2 x 10^{-6}$ at the photosphere of the sun, $\sim 3 x 10^{-5}$
in the solar system relative to the Great Attractor, $\sim 10^{-3}$
at the smallest AGN radii of which we can be certain, $\sim 0.17$ at
the innermost marginally stable orbit of a NS or BHC, $\sim 0.25$ at the
surface of a maximum mass NS in the SM and 0.5 at the
event horizon of a black hole. The five orders of magnitude jump from
the solar system testing ground to the neutron star regime is good reason for
caution about the acceptance of any strong-field theory of gravitation.

The exponential metric can be inferred
from special relativity and the principle of
equivalence applied to frames co-moving with an
accelerated particle (Einstein 1907, Rindler 1969, Yilmaz 1975).
This is a strong indication that the exponential metric is
the proper static limit for gravitation. That
this metric does not satisfy the field equations of
General Relativity was one of the motivations for the
earliest version of the Yilmaz theory (Yilmaz 1958). It
is noteworthy that the the exponential metric is merely
rescaled by addition of an arbitrary constant to the
potential whereas the SM depends on
an absolute potential. Without this dependence, event horizons could be
transformed away. With the conventional choice of
zero potential at infinity, however, the two metrics
are the same to first order in u(r). This is
sufficient to insure that they give the same results in
the four classic weak-field tests of General
Relativity. Gravitational redshift of
radiation observed distantly is given by (Rindler 1969):
\begin{equation}
z~=~\exp(u(r))~-~1
\end{equation} and in the SM:
\begin{equation}
z~=~\frac{1}{\sqrt{1~-~{2 u(r)}}}~-~1
\end{equation}

(i) Accretion Mechanics\\
In either metric there is an innermost marginally stable orbit that can
be reached by an accreting particle. The orbital radius is (see appendix)
$r_{ms} = 4/(3-\sqrt{5})GM/c^2$ in the YM.
Thus $u(r_{ms})=0.191$. In the SM $r_{ms} = 6GM/c^2$
and $u(r_{ms}) = 1/6$. An accreting particle starts from a large radial
distance, with essentially zero momentum and $u(r) \approx 0$.
Its energy is $E = m_0 c^2$. At $r_{ms}$ in the YM, the energy is
$E = 0.945 m_0 c^2$ (see appendix). The binding energy at $r_{ms}$
is the difference, $0.055 m_0c^2$. Whether this is radiated or advected
depends on the opacity of the accretion disk and the viscous
dissipation mechanisms.  For any circular orbit in
the disk, the particle energy
in the YM can be shown to be $E = m_0c^2 \exp(-u)\{(1 - u)/(1 -
2u)\}^{1/2}$. Thus, in the YM the maximum
fraction of accretion mass energy which can be observed
as radiation from the disk is: 
\begin{equation}
f_{d}~=~1~-~{\exp(-u)}\sqrt{\frac{1-u}{1-2 u}}
\end{equation} limited to u $<$ 0.191.
For the SM the maximum fraction of
accretion mass energy which can be radiated from the
disk, limited to u $<$ 1/6 and 0.057m$_0c^2$, is:
\begin{equation}
f_{d}~=~1~-~\frac{1~-~2 u}{\sqrt{1~-~3 u}}
\end{equation}
If the innermost marginally stable orbit is the inner
disk boundary, disk mechanics and disk luminosity
should be only weakly mass dependent and very nearly the
same for the two metrics. The binding energy at $r_{ms}$ differs
by only 3.6\% in the two metrics. It is
independent of both the mass of the central object and $r_{ms}$.
For the star model used here, disks would
terminate at r$_{ms}$ above the surface for NSs with masses above
2.1 M$_\odot$ (YM) or 1.7 M$_\odot$ (SM).
For stars with sufficiently weak magnetic fields ($< 10^7 G$) this
``gap accretion'' would produce a hard power-law spectrum (Kluzniak \& Wilson
1991, Hanawa 1991, Walker 1992) arising from bulk flow Comptonization
in a boundary layer. In fact, however, radiation (Miller, Lamb \& Psaltis 1997)
and/or magnetic torques (Lai 1998 ) on the disk
appear to cause the inner disk radius of NSs to be larger yet.
At present it seems likely that kHz QPOs may be generated at
the sonic point just inside the disk (Miller, Lamb \& Psaltis 1997).
The observed QPO frequencies indicate that the sonic radius is generally
larger than $r_{ms}$, but may approach $r_{ms}$ as a limit at sufficiently
high accretion rates (Zhang, et al. 1998).

The Keplerian frequency of the innermost marginally stable orbit
is of considerable interest. In the YM, Keplerian frequencies are given by:
\begin{equation}
\nu_K = \frac{\sqrt{GM} exp{(-2u)}}{2 \pi r^{3/2} \sqrt{(1 - u)}}
\end{equation}
For the innermost marginally stable orbit, u = 0.191 and $r_{ms} = 5.24GM/c^2$,
this yields $\nu_K = 2040 M_\odot / M$ Hz, compared to 2200 M$_\odot / M$ Hz
in the SM; a mere 7\% difference.

For neutron stars there is additional energy dissipated as accreting particles
are brought to rest on the surface. Neglecting the effects of star spin and
evaluating the energy-momentum
four-vector for zero momentum shows the energy of a
particle at rest on a star of radius R to be E = m$_0c^2\sqrt{g(R)}$.
Thus the total accretion energy, external to the star, which
may be directly radiated to distant observers is $m_0c^2 (1 - \sqrt{g(R)})$.
The fraction radiated from the star
surface itself (neglecting fusion, which can be
considered as part of the binding to the star) would
be:
\begin{equation}
f_{s}~=~1~-~\sqrt{g(R)}~-~f_{d}
\end{equation}\\
In the strong-field cases to be considered here, over
half of the energy release exterior to the star occurs
within 3R and $\approx$ 90\% from within 10R.

(ii) Compact Objects\\
A compact object of constant proper density (Clapp 1973) is considered here.
The details are straightforward, but tedious, and have been placed
in the appendix. The proper density is $\rho(r) g(r)$ = $\rho_0$, a constant.
These compact objects have a characteristic
radius and mass that are given by:
\begin{equation}
r_0 = \frac{c}{\sqrt{4 \pi G \rho_0}}
\end{equation}
\begin{equation}
M_0 = \frac{c^2r_0}{G}
\end{equation}
\begin{figure}
\plotone{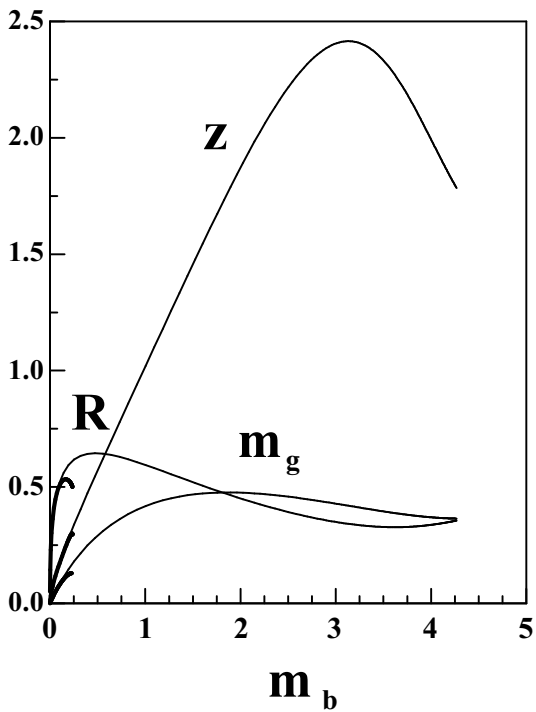}
\figcaption{ Radius, R, gravitational mass, m$_{ g}$, and
surface redshift, z, as functions of free
baryon mass, m$_{ b}$, contained within a neutron
star in the exponential metric. Radius is in
units of $r_0 = c/\sqrt{4 \pi G \rho_0}$
and masses in units of
$M_0 = c^2r_0 / G$. Heavy line features at the lower
left are the same quantities for the
SM.}
\end{figure}
As shown in Table 3, g(R) reaches 0.25 at the surface of a maximum
mass NS in the SM. Assuming that $\rho(R) = 2.7 x 10^{14}$ $g/cm^3$,
the nuclear saturation density, and using $g(R) = 1 - 2u(R)$ yields
$\rho_0 = 1.35 x 10^{14} g/cm^3$. This value of $\rho_0$
yields: $r_0 = 28.2$ km and $M_0 = 19 M_\odot$. In turn, these yield a maximum
mass of $2.45 M_\odot$ for 14.3 km radius and a maximum radius of 15.5 km for
the SM. Independent of the choice of $\rho_0$, NSs in the
upper one third of the mass range permitted by the
SM would have u(R) $>$ 1/6 and have radii small enough for
gap accretion. 

For the same $\rho_0$ in YM, the maximum mass is 9 M$_\odot$ at a radius
of 13.5 km. A maximum radius of 18.8 km occurs for about 5.1 M$_\odot$.
Mass vs.radius curves are similar in shape to those obtained by
Friedman \& Ipser (1987) for their core model. Although the constant proper
density model fails to properly account for the softer exterior layers
for low mass objects, this will be of no concern here. The effects of stellar
rotation may increase maximum masses (Friedman \& Ipser 1987) by about 25\%, to
about 3 M$_\odot$ for the SM and 11 M$_\odot$ for the YM.
Figure 1 shows comparisons of mass, radius and surface redshift for YM and SM.
It is interesting that there exists a maximum mass for the YM that is
dependent on density. For nuclear density, an 11 M$_\odot$ maximum mass
may be adequate to cover the range of masses that have been found for the
galactic BHCs. Figure 2 shows properties for the SM in more detail.

\begin{figure}
\plotone{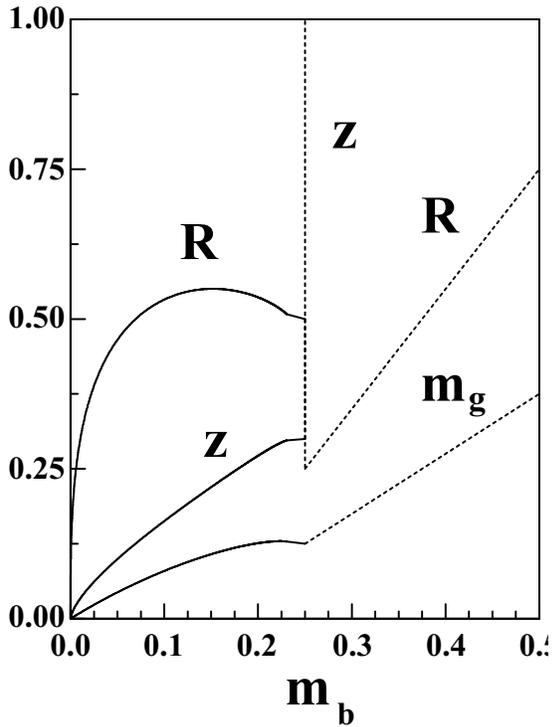}
\figcaption{ Radius, gravitational mass and surface
redshift as functions of free baryon mass for
a neutron star in the Schwarzschild metric.
Units are the same as Figure 1. Dashed lines
show extensions for black holes.}
\end{figure}
Tables 2 and 3 give masses, radii, surface potentials, surface redshifts, etc
for the two metrics. With these quantities, the necessary binding
fractions of the disk and surface can be calculated using
equations 7 - 9. Figure 3 shows these binding fractions.
A profound effect of a neutron star surface is
implied by Fig 3. Up to about $\approx$ 1.4 M$_\odot$, the luminosity
would be fairly evenly divided between disk and surface, but
for the largest masses in the Schwarzschild metric up
to 80\% of the luminosity might arise inside the last stable orbit;
near or on the star surface. In the YM as much as 90\% of the
luminosity could originate from the star surface. The binding fractions
at the star surface can reach 70\% for large masses in the YM. Large
binding fractions and redshifts would make their surface radiations
very prominent and very soft.

\begin{figure}
\plotone{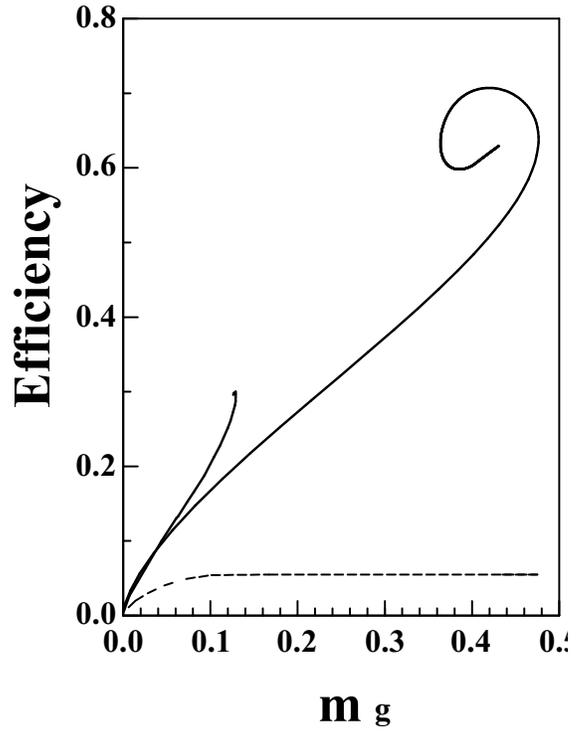}
\figcaption{Efficiencies of conversion of gravitational
potential energy to radiation as functions of
gravitational mass, $m_g$. Gravitational mass is
in units of $M_0$. Solid upper lines are $(f_s + f_d)$.
Free baryon mass, $m_b$, increases from left to
right along these curves. The spiral of the
YM occurs because maxima and
minima of redshift and gravitational mass do
not correspond to the same free baryon mass. 
Lower dotted line is $f_d$, which is essentially
the same in both metrics. Surface radiation
is represented by the difference between the
upper and lower lines, excepting black holes.}
\end{figure}

(iii) Internal Processes\\
After accreted matter reaches the surface there is
still energy to be produced by fusion and a need for
additional gravitational binding energy to be released
for stability of the star. In order to properly account for all of
the energy available, one needs to consider the
difference between free baryon mass and mass bound
gravitationally within the star. The total
gravitational energy release as rest mass $\Delta m_b$ falls
from a large distance and becomes bound in the star
would be $c^2 (\Delta m_b - \Delta m_g)$, or
$c^2 \Delta m_b (1 - dm_g/dm_b)$. The term in
parentheses being the maximum possible fraction of
accreted rest mass to be converted to radiations
seen by a distant observer. It is plotted in
Figure 4 as the gravitational binding fraction of the
object. For a neutron star in the SM,
it actually diverges shortly after it reaches 1.0 just
below 2.35 M$_\odot$. At this point the constant intrinsic
density model fails in the SM. The addition of more
mass simply causes a core collapse, though it may be possible that
some of the binding energy would escape through
neutrino emission, pair production and gravitational
radiation. 

\begin{figure}
\plotone{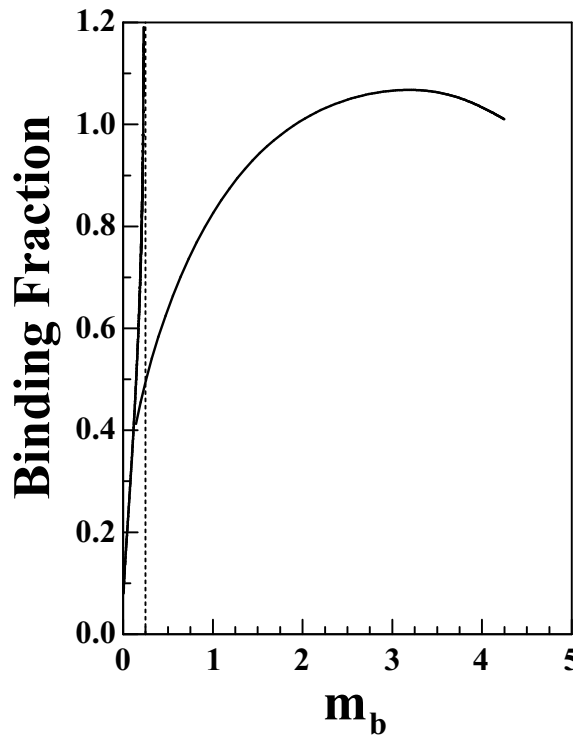}
\figcaption{ Binding energy expressed as a fraction of the
rest mass-energy of accreted particles; given
by ( 1 - $\frac{dm_g}{dm_b} )$. Collapse of the core in
the Schwarzschild metric is shown by the
diverging line at the left. The curved line
extending to the right is for the Yilmaz
metric.}
\end{figure}
In the YM there is no discontinuity of
binding fraction but after trapped baryons produce a
maximum gravitational mass, the star can become more tightly bound.
The more massive neutron stars permitted in either metric swallow
baryons but must radiate their mass equivalents for stability.
It is apparent that the overall binding fraction is much larger
than the binding fraction at the surface. Some substantial additional
radiations of internal origin should be seen from accreting NSs
in either metric. It is possible that this energy
would appear as copious photon, pair production or neutrino emissions,
or gravitational radiation. Pair plasma emissions might be part of the jets
observed for some NSs and BHCs (Gliozzi, Bodo, \& Ghisellini 1998).
Whether or not accretion energy might
be transiently stored has not been considered here, but it
should be noted that mass-energy can be trapped inside the photon sphere
for $u(R) > 1/2$ (YM) or 1/3 (SM). Spectral qualities of surface radiations
should also be affected when the surface lies inside the photon
sphere, however, this would involve analysis well beyond the scope
appropriate for this work.

\section{Magnetic Phenomena}
The magnetosphere radius can be estimated as the radius for which
the impact pressure of accreting matter is some fraction, $\eta$,
of magnetic pressure (e.g., Lamb, Pethick \& Pines 1973). For
dipole fields and disk accretion, with Keplerian motion outside the
magnetopause, this criterion yields :
\begin{equation}
r_m = {({\frac {\eta^2 \xi^2 B^4 R^{12}}{2 \epsilon^2 GM\dot{m}^2}})}^{\frac {1}{7}}
\end{equation}
where $\epsilon$ is the ratio of radial flow velocity to free fall velocity
in a spherical flow, $2\xi$ is the ratio of disk thickness to radius and B
is the magnetic field at R. A choice of $(\eta \xi / \epsilon)^{2/7} = 0.35$
brings the magnetospheric radius into close agreement with that of the much
more complicated model of Ghosh \& Lamb (1992).
Using $ \eta \xi / \epsilon = 0.025, M = 2.8 x 10^{33} M_{1.4} g,
B = 10^9 B_9 G, R = 10^6 R_6$ cm and $\dot{m} = 10^{15} \dot{m_{15}}$ g/s
this becomes:
\begin{equation}
r_m = (78 km){(\frac{ B_9^4 R_6^{12}}{M_{1.4} \dot{m_{15}}^2})}^\frac{1}{7}
\end{equation}
The magnetosphere radius shrinks to accomodate higher impact pressure.
It is relatively impermeable to diamagnetic plasma, but unstable for curvature
perturbations (Rayleigh-Taylor). Arons et al. (1984) have described shot
penetrations of the magnetopause and their further fragmentation via
Kelvin-Helmholtz processes as they fall through the co-rotating magnetosphere.
At relatively low accretion rates individual shots may produce considerable
flickering. The larger surface binding fractions for more massive objects in
the YM might explain the larger amplitude flickering observed for BHCs.
For recent reviews of magnetospheric phenomena, see Campana et al.
(1998a) and Becker and Tr\"{u}mper (1997).
It is well-known that pulsars can generate very hard spectra via
magnetospheric effects (Michel 1991, Becker \& Tr\"{u}mper 1997) and surface
impact (Alme \& Wilson 1973) as well as in
shock fronts where magnetospherically expelled matter impinges on nebulae.

The star spin frequency determines the co-rotation radius, $r_c$, for which
the magnetospheric equator rotates at the Keplerian orbit frequency.
In (SM), $r_c = {(GM / 4 \pi^2 \nu_S^2)}^{1/3}$. When the magnetospheric
radius lies outside the co-rotation radius, accreting matter receives a
super-Keplerian push at the magnetopause. It is swept
outward and is unable to penetrate the magnetosphere until a boundary
layer with sufficient pressure builds. This is the
well-known ``propeller effect'' (Illarianov \& Sunyaev 1975).
When the magnetosphere is forced inside the co-rotation radius, the spectrum
changes from a hard, low state to a state of higher luminosity and softer
spectrum due to greater surface contributions. If the transition radius is
large there may be a large flare as mass strikes the star. Since the co-rotation
radius is determined by the spin, there is a strong luminosity-spin correlation
for these flares (Stella, White \& Rosner 1986). Additional evidence has
recently been found (Cui 1997) for propeller effects in
the accreting x-ray pulsars, GX 1+4 and GRO J1744-28. Sharp transitions
were observed at luminosities of $\sim 10^{37}$ erg/s, for which
x-ray pulsations ceased, but became re-established at slightly
higher luminosity levels. Zhang, Yu \& Zhang (1998) and
Campana, et al. (Campana, S. et al. 1998b) have independently
analyzed the decline of the 1997 outburst of Aql X-1 and found strong
evidence of a propeller effect and an accompanying spectral state
transition with very little luminosity change. The transition was
followed by a rapid cutoff of luminosity that was arrested only briefly as
radiation pressure from the surface emissions ceased.

For sufficiently large $r_c$, the maximum luminosity that
can be achieved without surface contributions
corresponds to $r_m \approx r_c$ and $L_c = GM\dot{m}/2r_c$:
\begin{equation}
L_c=  \frac{2 \sqrt{2} \eta \xi \pi^3 B^2 R^6 \nu_S^3}{\epsilon GM}
\end{equation} or
\begin{equation}
L_c = (1.2 x 10^{34}erg/s) {B_9}^2 {R_6}^6 {\nu_2}^3 {M_{1.4}}^{-1}
\end{equation}where $\nu_S = 100 \nu_2$ Hz. Campana et al. (1998) have
produced the same equation for spherical accretion
except for a multiplier of 20 in place of
the 1.2 used here. The smaller multiplier merely reflects the
smaller magnetospheric area impacted by a disk.

For objects more massive than $\sim 1 M_\odot$, classical physics expressions
for surface luminosity are inadequate, however, we
can use $L = (y f_s + f_d) \dot{m}c^2$, where y is the fraction of
mass reaching the magnetopause that penetrates to the surface.
Using this to evaluate $\dot{m}$ and setting $r_m \approx r_c$ in Eq. 13
provides a relation of the luminosity for accretion at the co-rotation radius
to the magnetic field and spin of the star:
\begin{equation}
L = \frac{ {(2 \pi)}^{\frac{7}{3}} (y f_s + f_d) c^2 \xi \eta B^2 R^6
\nu_S^{\frac{7}{3}} } {\sqrt{2} \epsilon {(GM)}^{\frac{5}{3}} }
\end{equation} or
\begin{equation}
L = (9 x 10^{35} erg/s)(y f_s + f_d) B_9^2 R_6^6 \nu_2^{\frac{7}{3}}
M_{1.4}^\frac{-5}{3}
\end{equation}
Eqs. 16 and 18 should yield identical results for y = 0, however, since the
classical and relativistic binding energies differ slightly for the
inner disk, there can be spurious small differences.

$L_{min}$ will be used to designate the accretion rate for which y=1,
$r = r_c$, and all of the accreting matter reaches the star.
If the co-rotation radius is relatively large compared to the star radius,
large luminosity changes can occur
at the transition from $L_c$ to $L_{min}$, even if the accretion rate changes
very little. Even slight departures from y = 0 might produce such
strong radiation pressure that the full transition to $L_{min}$ is prevented.

The magnetosphere is limited by the light
cylinder radius, $r_{lc}$, at which a co-rotating
magnetosphere equator would move at the speed of light. For
fast spinners with $\nu_S \sim 400 Hz, r_{lc}$ can be less
than $\sim 120 km$. When the magnetospheric radius lies beyond $r_{lc}$
the radio pulsar mechanism may become active. Both pulsed point source
x-ray emissions and diffuse harder radiation at lower luminosity from a
synchrotron nebula have been observed for the isolated $\gamma$-ray pulsar
PSR 1055-52 (Shibata et al.1997 ApJ 483, 843).

The maximum quiescent luminosity that can be produced solely by
disk accretion corresponds to $r_m = r_{lc}$. With $L = GM \dot{m}/2r$, this
yields:
\begin{equation}
L_{q,max} = \frac{ \xi \eta {(2 \pi \nu_S)}^{\frac{9}{2}} \sqrt{GM}
B^2 R^6}{ 2 \sqrt{2} \epsilon c^{\frac{9}{2}}}
\end{equation}
or
\begin{equation}
L_{q,max} = (3.4 x 10^{30}erg/s) {M_{1.4}}^{\frac{1}{2}} {B_9}^2 {R_6}^6 {\nu_2}^{\frac{9}{2}}
\end{equation}
If a pulsar wind drives the quiescent luminosity, the rotational energy
loss rate is (Bhattacharya \& Srinivasan 1995):
\begin{equation}
\dot{E} = 4 \pi^2 I \nu_S \dot{\nu_S} = \frac{32 \pi^4 B^2 R^6 \nu_S^4}{
3 c^3}
\end{equation}
where I is the moment of inertia of the star. The right member of
this equation is empirical, lacking magnetic axis inclination dependence and
perhaps factors of $\eta$ and $\epsilon$, however, it and the middle member have
been taken as an operational definition of the magnetic moment, $B R^3$,
for NSs with $I = 10^{45} g-cm^2$. Becker and Tr\"{u}mper (1997) have
provided a tight correlation between x-ray luminosity and rotational
energy loss rate for which the x-ray luminosity is
$1.0x10^{-3} 4 \pi^2 I \nu_S \dot{\nu_S}$ again with fixed I = $10^{45}$
g $cm^2$. While most of the correlated sample likely consists of stars near
$1.4 M_\odot$, what has really been correlated is x-ray luminosity and
$\nu_S \dot{\nu_S}$. Had a different value of I been used
in the calculation of $\dot{E}$, the multiplier that would correlate
with the x-ray luminosity would be $1.0x10^{-3}x10^{45}/I$. For use with the YM,
$I = 2.85 x 10^{45}~g-cm^2$, appropriate for $1.4 M_\odot$ and $R_6 = 1.5$
will be used here. In this way the YM model can be satisfied while maintaining
relatively accurate x-ray luminosities. Combining this with the right member
of Eq. 21 yields:
\begin{equation}
L_q = (1.3 x 10^{30}~ erg/s) B_9^2 R_6^6 \nu_2^4
\end{equation}
In contrast to previous equations, this is merely a well grounded
empricicism that represents some loosely defined x-ray luminosity rather
than a bolometric luminosity. Still it is to be preferred over $L_{q,max}$ for
quiescent luminosity.

For accretion to reach the star surface, with quiescent luminosities
of $10^{31} - 10^{33}$ erg/s being above $L_c$, would require a combination
of very weak field and small co-rotation radius (fast spin). If most NSs
reach a spin equilibrium for long time average luminosities near those
in accord with Eq. 18 with y = 1, (White \& Zhang 1997) then their
quiescent luminosities are highly unlikely to
derive from either surface accretion or accretion at the light cylinder
radius.

Becker and Tr\"{u}mper's (1997) strong correlation of x-ray luminosity
with magnetic field strength and spin rate down to luminosities well below
$10^{30}$ erg/s strongly suggests that a preponderance of the quiescent
luminosity is magnetospheric. There are reasons for believing that the
power-law portions of quiescent emissions originate near the magnetopause
(see below). Nevertheless, some quiescent luminosity may derive
from a cooling NS surface. Excess soft radiation corresponding to
$kT_{bb} \approx 0.1 - 0.2$ keV and corresponding
to small $\sim 10 - 100 km^2$ areas has been found to be a part
of the quiescent emissions of both NSs and BHCs (Tanaka \& Shibazaki 1996).
It is conceivable that these could be polar cap emissions of NSs, but
Heindl \& Smith (1998) have shown that proper attention to flourescence and
reflection substantially reduces the apparent temperature of the excess
soft emissions. Further, even the remaining apparent temperature may
be larger than the surface temperature of a NS, depending on its atmosphere.
As shown in dips, some of the soft excess also appears to originate from an
extended disk as well. Thus the origins and magnitudes of contributions
to the quiescent luminosity are somewhat unclear at the present time.
For this work, for luminosities generally above $10^{32}$ erg/s, it will be
assumed that the power-law magnetospheric emissions dominate the hard states.
Ultrasoft blackbody ($kT_{bb} = 0.048$ keV) surface
radiation has been observed for the isolated gamma-ray pulsar, Geminga
(Halpern \& Wang 1997). No evidence of polar cap heating was found.
Non-thermal power-law magnetospheric emissions in the spectral range 0.5 to
2 keV were present. Obviously, ultrasoft and power-law spectral
components are not signatures of black holes. Geminga is a
high energy gamma-ray source with a $1.6 x 10^{12} G$ magnetic field. It is
the only known radio-quiet, rotation-powered pulsar with a strong
magnetic field, however, some of the BHCs might also have strong
fields; see below.

Brown, Bildsten \& Rutledge (1998) have found that crustal heating during
accretion episodes resets the core and surface temperatures of NSs and might
allow them to produce quiescent luminosities of as much as
$\sim 10^{32 - 33}$ erg/s. An actual temperature of as little as
$\sim 0.03$ keV, would contribute $\sim 10^{31}$
erg/s to the quiescent luminosity. This would produce a radiant equilibrium
temperature of about 5000 K for optically thick matter at
a distance of about $10^5$ km. The inner accretion disk would need to
be beyond this for thermal-viscous stability. For spins
in excess of 0.005 Hz, the light cylinder radius is smaller than
$10^5$ km. This effectively eliminates $L_{q,max}$ for the production of
quiescent luminosity. It also sets a minimum inner radius
at the start of outbursts. The inner disk must fill on a viscous
timescale, producing a delay of days between the onset of
optical and x-ray luminosities. Thus the 6 day
delay (Orosz, et al. 1997) between the onset of optical and x-ray
luminosity increases for the 1996 outburst of GRO J1655-40 would be expected.
Hameury, et al (1997) have calculated 6 days as the interval for viscous
inflow from $2x10^5 km$. Delays for refill of the
disk have been observed for dwarf novae and for the NS, 4U 1608-52.

\section{The Magnetic Fields of NSs and BHCs}
As additional information is acquired, some of the tentative conclusions
drawn here will surely need alteration. The examples chosen are those for
which I have sufficient information to initiate a discussion. They
do not all fit neatly into any current scheme.

(i) Neutron Stars\\
It is well established that radio pulsars may have magnetic fields ranging
from $10^{8}$ G for millisecond pulsars to $10^{13}$ G for slower spins.
If not for spins revealed by bursts or kHz QPO frequency differences, it would
be difficult to determine the magnetic fields of Atolls and Zs. They are not
pulsars in any normal sense. Nevertheless, ratios of luminosities can be
used to determine their spins and magnetic moments. Since all
of the luminosity equations considered here depend on  the
dipole moment, $B R^3$, it can be eliminated by the use of luminosity ratios,
thus permitting the spin to be estimated. Once the spin is known, B can be
calculated. For consistency with the constant
proper density star model, R = 15 km for a 1.4 M$_\odot$  will be used here.
Since the quantities being determined from luminosity ratios are actually
spins and dipole moments, an overestimate of R may cause an underestimate of
B without other effect. In order to use the ratio method, two of $L_{min}$,
$L_c$, $L_{q,max}$ or $L_q$ must be identified.

\textit{Aql X-1} is an important test case for the use of ratio methods.
Campana et al (1998b, Fig. 1) reported spectral hardening beginning at
$L_{min} =  1.2x10^{36}$ erg/s and complete cessation of the rapid
decline at about $1.2x10^{33}$ erg/s. Identifying these as $L_{min}$
and $L_{q,max}$, the ratio yields $\nu_S = 512$ Hz for $m_{1.4} = 1$.
This agrees moderately well with a spin frequency of 549 Hz observed
during burst QPOs. Using the QPO spin frequency and Eq. 18, a
magnetic field of $1.3x10^8$ G can be calculated.
Aql X-1 displays ultrasoft and power-law components in its quiescent
spectrum with the power-law component accounting for more than half the
luminosity. Using 549 Hz and $1.3x10^8$ G, the magnetospheric component
of quiescent luminosity is calculated to be $2.3x10^{32}$ erg/s.
Garcia, McClintock \& Narayan (1997)
report a quiescent luminosity of $4 x10^{32}$ erg/s. Further, using 549 Hz and
$1.3 x 10^8$ G permits the calculation of $L_c = 3.8 x 10^{35}$ erg/s,
as obtained by Campana et al (1998).

\textit{SAX J1808.4-3658}
A second test case is provided by the recent outburst of the 401 Hz
x-ray pulsar and burst source, SAX J1808.4-3658, (Gilfanov et al. 1998,
Heindl \& Smith 1998) which displayed a light curve very similar to Aql X-1.
It reached a luminosity level of $5 x 10^{36}$ erg/s, declined slowly to about
$1.5 x 10^{36}$ erg/s and then declined rapidly with a brief arrest at
$4 x 10^{35}$ erg/s before reaching about $10^{33}$ erg/s. Between the start
of rapid decline and the brief arrest, 401 Hz pulsations changed from 4\% rms
to undetectable, thus indicating that surface impacts ceased. Identifying the
luminosities at the start of rapid decline and brief arrest as $L_{min}$ and
$L_c$ allows the calculation of a magnetic field strength of $2.1 x 10^8$ G,
a convincing spin frequency of 419 Hz, $L_{q,max} = 8 x 10^{32}$ erg/s, and
$L_q = 1.7 x 10^{32}$ erg/s.

When the accretion rate is declining, the onset of the propeller effect
may be followed by a rapid cutoff of luminosity.
Although many NSs (and BHCs) have shown a steep decline of luminosity
from the vicinity of $10^{36}$ erg/s, this may not always be clear evidence
of the onset of a propeller effect. Without the cessation of pulses it would
have been unconvincing to identify the propeller effect for SAX J1808.4-3658.
Observations of Type I bursts at luminosities above $10^{36}$ erg/s show
that accreting mass reached the surface. Yet the spectrum remained
a hard power-law extending to 100 keV over two decades of luminosity change.
A detailed spectral analysis (Heindl \& Smith 1998)
revealed the presence of iron flourescence and a small, not very
well constrained, soft excess which was modeled as a MBB. An additional
unusual feature of this source is the soft x-ray lags reported by Cui,
Morgan and Titarchuk (1998). An obvious possibility for explaining the
unusual spectral and timing characteristics would be a very small inclination
view with strong polar axial magnetospheric emissions obscuring the
surface. Alternatively, there might be some unusual surface mechanism that
produced the hard spectrum. At any rate, the presence of only hard spectral
components cannot be interpreted as a lack of surface accretion.

The spin frequencies determined from luminosity measurements
for Aql X-1 and SAX J1808.4-3658 are in convincing agreement with burst QPO
frequencies, however, if their assumed masses had been doubled, their
calculated spins would have been halved and their magnetic fields roughly
doubled. For 4U 1636-54 and KS 1731-26, the frequency difference of kHz QPOs
has been near half of the burst QPO frequency (Swank 1998). The question arises
as to what is the true spin frequency. In this case it seems to be the
burst QPO frequency. It has been suggested that the burst QPO frequency
might be double the actual spin frequency due to two antipodal hot spots on
the star. It seems improbable that bursts would be simultaneously
initiated at two different places on the star surface (Strohmayer et al. 1998).
On the other hand, the detonation of some $10^{18}$ megatons at one pole might
quickly ignite another pole less than 50 km distant, depending on the shielding
afforded by an intervening stellar mass (Miller 1998).

\textit{4U1608-52} provides another test case for the use of ratios.
Spectral hardening began in decline at $L_{min} = 10^{37}$ erg/s (Mitsuda et
al. 1989). Garcia, McClintock \& Narayan have reported a quiescent luminosity
of $L_q = 2x10^{33}$ erg/s, while Tanaka \& Shibazaki (1996) report $6x10^{32}$
erg/s.  For these luminosities, the ratios $L_{min} / L_q$ and
Eqs. 18 and 22 yield spins of 566 Hz and 275 Hz. The frequency difference
of twin kHz QPOs for this source is near 230 Hz, however, this might be only
half of a 460 Hz actual spin frequency. The conflicting luminosity data leave
this case unresolved, but it illustrates an important point; that the results
are probably correct to within small numerical factors anyway.
The magnetic fields estimated from $L_{min}$ and the two QPO spins are
$4.6x10^8$ G and $8.4x10^8$ G. The higher spin and weaker field are
the more likely attributes of an Atoll source.

\textit{4U 1820-30} apparently has a mass in excess of 2 M$_\odot$ based on
its kHz QPO limit frequency of 1060 Hz (Zhang, et al. 1998). Based on QPO
frequency differences it spins at either 275 or 550 Hz. Both of the high
frequency QPOs were visible in the low state at a luminosity of
$1.5 x 10^{37}$ erg/s but cut off when the spectrum softened
for luminosities in excess of $3.1x10^{37}$ erg/s. Assuming that the start of
twin QPOs would correspond to mass reaching the surface implies $L_{min} =
1.5 x 10^{37}$. For 2 M$_\odot$, Eq 18 yields either $5x10^8$ G
or $1.1x10^9$ G. The higher field would be an unusually strong field
for an Atoll source, but perhaps not for one of 2 M$_\odot$.
Calculated quiescent luminosities for the two cases are
$5x10^{33}$ erg/s and $1.5x10^{33}$ erg/s.
Spin and magnetic field estimates for other Atoll
sources, 4U 2129 + 47, KS 1731-26 and 4U 1730 - 335 are included in Table 1.

\textit{Cir X-1}: Neither spin rate nor mass are available for the enigmatic
Cir X-1. $L_{\min} = 6.3 x 10^{38}$ erg/s and $L_c = 3 x 10^{37}$ erg/s
can be determined from observations just before and two hours after
a spectral hardening transition on Sept 20-21, 1977
(Dower, Bradt \& E Morgan 1982). The ratio permits an estimate
of a 32 Hz spin followed by a magnetic field estimate of $8 x 10^{10}$ G.
These results were obtained for $M_{1.4} = 1.0$ and $R_6 = 1.5$. If the mass
of Cir X-1 were $7 M_\odot$, the spin would be 38 Hz and the magnetic field
$8.2 x 10^{10}$ G. The luminosity ratio method is relatively insensitive to
mass, provided that the correct accretion efficiency is used.
The calculated quiescent luminosity from Eq. 22 is $(0.9-6)x 10^{33}$ erg/s,
which is well below detector sensitivities.
Cir X-1 experiences a much wider range of accretion rates on orbital
time scales than most NSs. Bradt, Shirey and Levine (1998)
have provided a fascinating series of hardness-intensity diagrams
for this source.

The Z and Atoll NSs of Table 1 have relatively small differences
between $L_{min}$ and $L_c$ due to their rapid spins. The ratio
of these, $\Delta$, should be:
\begin{equation}
\Delta = \frac{y f_s + f_d}{f_d}
\end{equation}\\
To first order with y = 1, $r = r_{ms}$ in the numerator
and $r = r_c$ in the denominator, $\Delta \approx 2u(R)/u(r_c) \approx 2r_c/R
\approx 2 \sim 3$ for fast spins. This result has been discussed for
spherical accretion by Corbett (1996).
The bolometric luminosity change when the magnetopause reaches the
co-rotation radius may be much smaller than implied by the ratio $2r_c/R$
if $ y < 1$. 

If NSs of the fast spin Atoll and Z classes have been spun up to high
equilibrium spin rates by accretion, then the long-term average luminosity
would be near $L_{min}$. White and Zhang (1997)
used these luminosities and the known spins of several Atoll and Z class
NSs to estimate their magnetic fields. Values of $L_{min}$ or $L_c$
found for the spectral state transitions of several other
NSs have been used here to determine some of the NS magnetic
fields shown in Table 1. The results, though intentionally sparse,
fit the pattern suggested by van der Klis (1994) and found by White and
Zhang (1997). Z sources appear to have fields that are stronger than those
of Atolls. There appears to be a $L_{min} ~\alpha~ B^2$ correlation in
the results. White and Zhang suggested that the correlation might be spurious;
that there might be comparable magnetic fields in all cases with a
magnetospheric radius that is more weakly dependent on accretion rate
for Zs where radiation pressure dominates the inner disk. The clear
differences of properties of Atolls, Zs and Cir X-1 suggests that the
correlation represents a progresssion of a real physical property.

Kilohertz QPOs have now been observed for many NSs, including both
Atoll and Z types. The Atoll source 4U 1820-30 showed a limit frequency
of 1060 Hz, which appears to correspond to the innermost marginally
stable orbit for $\sim 2M_\odot$ (Zhang, et al. 1998).
This mass estimate obtained from the limit frequency of the QPO
must be taken seriously. Cyg X-2, a Z source with a mass $>1.9 M_\odot$
found from orbit parameters (Casares, Charles
\& Kuulkers 1998) has produced kHz QPOs of 1007 Hz but without reaching a
limiting frequency; perhaps implying a mass near 2M$_\odot$.
The BHC GRO J1655-40, at 7 M$_\odot$ (Bailyn et al.
1995) has produced a 298 Hz QPO while
in a high luminosity state (Remillard et al 1997). If this is the Keplerian
frequency of the marginally stable orbit it implies a mass of 7.4 M$_\odot$
(SM) or 6.8 M$_\odot$ (YM). One clear implication of these
results is that the accretion disks of both NSs and BHCs may reach
the innermost marginally stable orbit. With binding fractions limited
to about 6\% in all such cases, the differences between NSs and black
holes must be found inside the last stable orbit, in the differences
between surfaces and event horizons. The disks, excepting differences
due to irradiation from the central source, should be much the same,
(excepting those of Kerr metric black holes).

If kHz QPOs arise from clumps of mass falling from the inner disk radius
to the surface, then the disk inner radius must be less than the co-rotation
radius. Radii of Keplerian orbits at the frequencies at which kHz QPOs
first appear are generally about two thirds of the magnetospheric radii
determined from Eq. 13. Dong Lai (1998) has suggested that the sonic radius
might be identical with the magnetosphere radius when outside the innermost
marginally stable orbit. This might be accommodated here by suitable choice of
the impact pressure-magnetic pressure balance. To do so would reduce the
magnetosphere radii calculated from Eq. 14 and change the calculated dipole
moments determined from the luminosity equations without other effect.

(ii) Black Hole Candidates\\
If BHCs are really massive NSs that show neither pulses nor bursts
then spectral state switches and ratio methods must be used to
determine their spins and magnetic fields. As is the case for Atolls
and Zs, coincident spin and magnetic axes are necessary to account for the
absence of x-ray and radio pulses. For BHCs, it is possible in
some cases to treat the inner disk radius determined from MBB fits to
spectra as a co-rotation radius. If the massive NS hypothesis is correct, then
the substantial surface contributions may explain the small inner radii
that are often found for MBB fits to BHC high state spectra.
For low states, however, in which surface accretion
is largely blocked by propeller effects, the BHC spectral models are
quite applicable to NSs. Table 1 lists estimated spin and magnetic
field parameters for the BHCs discussed below.

\textit{ GRS 1124-68} shows characteristics that clearly illustrate
the nature of the spectral state switch. Misra \& Melia (1997)
showed that its inner disk radius increased to about $27 R_g$ ($R_g = 2GM/c^2$,
400 km for 5 M$_\odot$) after the spectral state transition.
Ebisawa, et al (1994) give a luminosity of $L_c = 6.6 x 10^{36}$ erg/s for
the low state. The assumption of $r_c = 400$ km permits estimates of
$\nu_S = 15$ Hz and $B = 1.1 x 10^{11}$ G. In turn $L_{min} = 2.4 x 10^{38}$
erg/s. Eq 22 yields $L_q = 3.9 x 10^{32}$ erg/s.  The calculated
$L_{min}$ fortuitously agrees exactly with observations. The
observational upper limit for quiescent luminosity is $4 x 10^{32}$ erg/s,
which is consistent with $L_q$ calculated for the pulsar wind mechanism.

$\dot{Z}$ycki, Done and Smith (1998)
fitted the spectrum of GRS 1124-68 with a power-law plus a reprocessed fraction
that included an iron line, and a relatively cool disk extending from
$r_{in}$ to  $5 x 10^4 R_g$ (with $R_g$ as defined above). A soft
thermal component was modeled separately by optically
thick Comptonization of soft seed photons from a MBB. By separately accounting
for soft components their model would not have $r_{in}$ forced to a small value
corresponding to a star surface, if one existed. They found little,
if any, change of $r_{in}$ at the transition to the intermediate spectral state.
There was a dramatic decrease of the degree of
ionization of the disk which they attributed to the disappearance
of the soft component. In addition, they attributed a
hardening of the power-law component to the decreased availability
of soft photons for Compton cooling. A strong reflected component
was found all the while the low state decline continued, which indicates
that optically thick material was present within about 700 km throughout
the observations that were analyzed. In all respects, these results exactly
accord with expectations for a massive NS.

\textit{ GS 2000 + 25}: $\dot{Z}$ycki, Done and Smith (1997a) have found a
similar spectral state switch for GS 2000 + 25. When a strong soft component is
present the disk is highly ionized and iron flourescence
strongly relativistically smeared. The disappearance of the soft component
is accompanied by hardening of the power-law and a decrease of ionization.
Little change of inner disk radius occurs at the transition, but the
subsequent decline of luminosity is accompanied by an increase of inner
disk radius. Using an inner radius of (50 - 100) $R_g$ (1040 - 2080 km
with $R_g = 2GM/c^2$) as the co-rotation
radius (Done, C. personal communication) and $L_c = 1.5 x 10^{35}$
erg/s for the hard state of Dec. 16, 1989, $\nu_S = (4.5 - 1.6)$ Hz
and $B = (1.4 - 6.5)x 10^{11}$ G are obtained. The calculated
quiescent luminosity is $(3.7 - 1.3) x 10^{30}$ erg/s; well below the
observational upper limit of $2 x 10^{32}$ erg/s.

\textit{ GS 2023 + 338} never displayed an ultrasoft spectral component,
but on May 30, 1989 it changed luminosity by a factor of 21 in 10 s,
from a saturated $10^{39}$ erg/s to $4.8x10^{37}$ erg/s (Tanaka \& Lewin
1995). $10^{39}$ erg/s is the Eddington
limit for 7 $M_\odot$. The iron flourescence line was not obviously
present in the high state but showed up prominently at the lower
luminosity. This might be caused by a distended inner disk covering the
magnetosphere. When the luminosity diminished, the reduction in the
1 - 10 keV band was greater than for higher energies. These phenomena
strongly suggest that the magnetosphere expanded beyond the co-rotation
radius when the luminosity dropped. Assuming this to be the case, apparently
$\Delta \approx 21$. For $7 M_\odot$ and $\Delta = 21 ,~u(R) = 0.56,~
f_s + f_d(r_{ms}) = 0.428$ (Table 2), and $f_d(r) = 0.02$. Solving for u(r)
yields 0.0434, which then gives r = 240 km as the co-rotation radius.
Using $\dot{m} = L/f_d c^2$ with L = $4.8x10^{37}$ erg/s gives
$\dot{m} = 2.7 x 10^{18}$ g/s. Solving Eq. 14 for the magnetic field then
yields $9.5x10^{10}$ G. The spin frequency (YM) determined from the
co-rotation radius is 39 Hz.

From spectral analysis $\dot{Z}$ycki, Done and Smith (1997b)
found $r_{in} = 263$ km and 368 km, respectively, for June 20, 1989 and
July 19-20, 1989. Luminosities (0.1 - 300 keV) for these dates
based on spectra of Tanaka (1992) are $1.9x10^{37}$ erg/s and $6.3 x 10^{36}$
erg/s, respectively. These radii give values of $f_d$ of 0.018 and 0.0134
and values of $\dot{m}$ are then $1.2x10^{18}$ g/s and $5.2x10^{17}$ g/s.
Values of B = 7.3 $x 10^{10}$ G and $8.8 x 10^{10}$ G then follow
from Eq. 14. The three independent estimates of magnetic field are about as
consistent as can be expected with an Eddington limit affecting the first
result. Stated another way, if the $9.5x10^{10}$ G field is used with the values
of $\dot{m}$ to calculate the magnetospheric radii, values of 286 km and 365 km
are found. These are within 9\% of spectral fit values that have
statistical uncertainties in excess of 50\%.

Using the 39 Hz spin, the quiescent luminosity would be $(5.2-8.9)x10^{33}$ erg/s
for $(7.3-9.5)x10^{10}$G. Quiescent luminosities of $6.9x10^{33}$ erg/s
Chen, Shrader \& Livio (1997), $8x10^{33}$ erg/s (Tanaka \& Shibazaki 1996),
and $1.6x10^{33}$ erg/s (Garcia, McClintock and Narayan 1997)
have been reported. In any event, the calculated
quiescent luminosity depends very strongly on spin, which is quite
uncertain in this case with $L_{min}$ so near the Eddington limit.

\textit{ 1E1740.7-2942} is a jet source. Vilhu et al. (1997)
fitted a "sombrero" model to a low state spectrum and found $r_{in} \sim
7 - 8 R_g$ for the $10 M_\odot$ they assumed. Taking this to be an inner
radius of $r_c = 225$ km and $L_c$ to be the corresponding
$3 x 10^{37}$ erg/s for $7 M_\odot$ yields $\nu_S > 43$ Hz
and $B > 6.9 x 10^{10}$ G. The hard spectrum with a small inner radius
is incompatible with the advective accretion flow (ADAF)
model of Narayan, Garcia and McClintock (1997). The calculated
quiescent luminosity is $<7x10^{33}$ erg/s.

\textit{ Cygnus X-1} has been observed to emit gamma rays (Ling et al. 1987).
It exhibits spectral state switches with very little
change of bolometric luminosity at about $5 x 10^{37}$ erg/s
(Belloni et al. 1996).  Misra \& Melia (1997)
obtained an inner disk radius of 56.7 $R_g$
from spectral fitting of an intermediate (hard) state.
For 10 M$_\odot$, this is 1700 km. Assuming this to be the co-rotation
radius yields a spin frequency of 2.6 Hz. Using R = 14 km in Eq 18 then yields
$B = 1.5 x 10^{13}$ G. Magnetic fields of this magnitude have been
associated with other gamma emitters such as the Crab and Vela pulsars.
A lack of pulses would require nearly coincident rotation and magnetic axes
and a lack of multipole moments. These are stringent, but not
impossible, requirements. The calculated quiescent luminosity, should
it ever be observed for comparison, is $1 x 10^{33}$ erg/s.

\textit{A0620 00}: Spectral hardening began about 100 days after the start of
the 1975 outburst and continued until interrupted by a reflare.
$L_{min} = 8 x 10^{36}$ erg/s (E. Kuulkers 1998) and
$L_q = 10^{31}$ erg/s (Garcia, Narayan \& McClintock 1997) along with Eqs
18 and 22 permit the estimates, $\nu_S = 13.5$ Hz and B = $2.3 x 10^{10}$ G.
The kinetic energy that the magnetosphere can impart to impinging matter
is generally much less than the binding energy for the inner disk,
but not for the light cylinder. This may be involved in the reflare process.
In the case of Atolls and Zs, the spins are so fast that matter is too
tightly bound at the light cylinder, hence little reflare is observed for them.

\textit{ GRO J1655-40}: $L_{min}$ was reached at $3.1 x 10^{37}$ erg/s
approximately July 29, 1996 (Mendez, Belloni and van der Klis 1998).
The rapid decline was arrested at $1.5 x 10^{36}$ erg/s, which can be taken
to be $L_C$. Eqs. 16 and 18 then yield $\nu_S = 34$ Hz and
$B = 2.1 x 10^{10}$ G. The calculated quiescent luminosity for these values
exactly matches the observed $2.5 x 10^{32}$ erg/s reported by Garcia,
McClintock \& Narayan (1997).

\textit{ GRO J0422 + 32} never showed an ultrasoft peak and stayed in
a spectrally hard, gamma emitting state during its 1992
outburst (Grove, et al. 1998). Lacking evidence of any spectral state
change, only possible limits for spin and field can be estimated here.
Using $3.6 M_\odot$, (Chen,Shrader \& Livio 1997)
maximum luminosity $7.9 x 10^{37}$ erg/s as $L_{min}$ and quiescent
luminosity $< 7.9 x 10^{31}$ erg/s (Garcia, McClintock and Narayan 1997)
as limits, one obtains $\nu_S < 0.36$ Hz and $B > 1 x 10^{14}$ G. Such large
magnetic fields may occur in nature, but it is possible that the spectrum
might have remained hard for the same reasons, whatever they were, as
SAX J1808.4-3658. A slow spin would be consistent with the
remarkable lack of high frequency features
in the power density spectrum. This lack is at odds with accretion models
that produce all hard x-rays and gamma emissions from regions near
the innermost marginally stable orbit (Grove et al. 1998). Grove et al.
found a strong QPO, independent of photon energy
or luminosity at 0.23 Hz. If this is the spin frequency, the
magnetic field could be as high as $2 x 10^{14}$ G.
Another possibility is that the rapid decay after about $10^{36}$ erg/s
might correspond to the onset of the propeller effect. This would
still imply a large magnetic field of $10^{12}$ G. In either of these slow
spin cases, the quiescent luminosity would be below $5 x 10^{31}$ erg/s;
still below the observational upper limit.

\textit{GRS 1915 + 105} displays oscillations with peaks above the
Eddington limit followed by hard states that are lower in luminosity
by a factor of 3 - 6. The conventional view is
that the inner disk becomes unstable and a flare is
produced by unspecified means as it
falls into a black hole. Belloni et al. (1997) have attributed the intervals
between flares to the time required for the inner disk to refill on
a viscous time scale, which is likely correct.
There is, however, a more credible explanation of the flares. They
could arise from a star surface. If plasma falling from the innermost
marginally stable orbit can produce a flare on
the surface that is beyond the Eddington limit, radiation pressure could
then push the disk back beyond this orbit and temporarily cut off the flow
to the surface. After the inner disk refills, another flare follows, etc.
MBB spectral fits (Belloni et al 1997) show that $r_{in}$ oscillates between
about 20 km and 80 km, but occasionally reaches only 55 km, followed immediately
by another burst. 55 km would be the innermost marginally stable
orbit radius (YM) for $7 M_\odot$, which is therefore adopted as the
mass.

A 67 Hz QPO has been observed (Remillard et al 1997,
Remillard \& Morgan 1998) that is sharp ($Q > 20$), stable
for factors of 5 luminosity change and stable over six months time.
It is observed most clearly for luminosities for which surface radiations
should be seen, if there is a surface. Speculating that this is the spin
frequency, a co-rotation radius of 163 km (YM) is implied.
The speculative 67 Hz spin frequency must be tested by further observations.
There are two sources of drifts and broadening for this QPO. Orbital
characteristics may imprint cyclical variations on it. Observing it at
high luminosities with radial expansion of the surface layers may broaden it.
If some of the hard secondary bursts that have been observed
(Taam, Chen and Swank 1997) originate on the
surface, as seems to be implied by the small radii determined for them,
they may also show identifiable spin QPOs.

The low state between flares reached about $4x10^{38}$ erg/s for a radius of
about 80 km. Evaluating $f_d$ for 80 and 163 km, calculating $\dot{m} =
9.3 x 10^{18}$ g/s for the 80 km radius and scaling it for 163 km,
($\alpha ~r_m^\frac{-7}{2}$) permits the calculation of $L_c = 1.9 x 10^{37}$
erg/s. If the accretion rate of $9.3x10^{18}$ g/s for the cutoff state were to
impact the surface after refilling the inner disk, the luminosity could be
driven to $3.6 x 10^{39}$ erg/s, if not for the Eddington limit
($\sim 10^{39}$ erg/s). The excess energy above the observed $1.6 x 10^{39}$
erg/s may well be what drives the jets. In the 10 - 50 s durations of
flares, excess energy of $10^{40} - 10^{41}$ erg might be imparted to jets.
The magnetic field, estimated from $L_c$ and Eq. 16 is
$2.8 x 10^{10}$ G. It is noteworthy that this exceeds the minimum $10^8$ G
found by Gliozzi, Bodo \& Ghisellini (1998) to be necessary
for the field strength at the base of the jets of GRS 1915+105.

Eq 22 yields an expected
quiescent x-ray luminosity of $7x10^{33}$ erg/s. In all parameters,
GRS 1915 + 105 seems to be like GRO J1655-40 and Cir X-1, both of which
are also chaotically variable in both x-ray and radio regimes.
It should be clear from the low value of $L_c$, that accreting mass should
have reached the surface at all times for the observed oscillation luminosities
under consideration. That it apparently did not, judging by the hard spectrum
after a burst, is due to the empty inner disk after the burst.
The star behaves as a relaxation oscillator between the super-Eddington
flares and innermost marginally stable orbit. For oscillations, it is critical
that the magnetic field be of the right strength to just require flows near
the Eddington limit to drive the magnetosphere near the marginally stable
orbit. Massive stars with very large surface efficiencies can
then provide a strong burst as mass hits the surface. The oscillating state
might require a balance of fields and accretion rates too
delicate for achievement by low mass neutron stars that have no accretion gaps.

\section{Spectra of Massive NSs}
For this discussion it must be remembered that there are only two
reliable distinguishing spectral characteristics of BHCs. These are
the simultaneous presence of hard spectral tails and ultrasoft spectral
components above $10^{37}$ erg/s and larger amplitude flickering for some BHCs.
Both are indicators of larger masses for BHCs in the massive NS model. No
event horizon explanation has been proposed for them for black hole models.
Flickering can be explained by shot penetrations of the
magnetopause, as illustrated by the flickering pulsar, VO332+53.
A shot mechanism has the virtue of being able to account for the strong
coherence (Nowak et al. 1998) of soft and hard emissions during flickering.
Stronger flickering of the BHCs is due to their larger surface binding energies.
The quenching of the hard spectral tail of the low mass NSs is due to the
fact that $10^{37}$ erg/s is a much larger fraction of the Eddington limit
for low mass NSs. At this and higher luminosities, radiation pressure inflates
the inner disk and the flood of soft photons cools the Comptonization
region. The magnetosphere would not be inside the co-rotation radius for many
BHCs at $10^{37}$ erg/s. These would automatically have hard spectra at this
luminosity.

Three separate regions contribute to the spectra of NSs. These are
the surface and co-rotating magnetosphere, the inner disk-magnetopause, and
the outer disk. It is accepted that blackbody and/or thermal
brehmsstrahlung spectra of $kT_{bb} \sim$ 2 kev arises from the region
near the surface. Hard power-law radiations seem to come from the inner
disk-magnetopause region, and soft excess themal radiations $kT_{bb} \sim$
0.1 - 0.6 kev originate from the outer disk. Not only are these
spectrally different regions, they have been spatially resolved in dip
observations. Morley et al. (1998) have analyzed deep (100\%) dips
in the high state spectrum of XB 1916-053 to show that $kT_{bb} = 1.95$ keV
blackbody emissions arise from a rapidly covered central NS
while a strong power-law spectrum beyond 100 kev with photon index of 1.75
was produced by a larger emission region. Observations of hard, low states
show dominant power-law spectra along with a low temperature soft excess
that produces a few percent of the luminosity.

Exactly the same spectral features and some of the same spatial features
have been observed for BHCs. For BHCs, however, the strong soft
radiations are attributed to the disk and are usually represented by
MBB radiation functions. It is a stroke of good fortune that MBB functions
represent the central soft sources of both NSs and BHCs, but certainly
no proof of an event horizon. A similar analysis of high state dips of the
BHC GRO J1655-40 (Kuulkers et al. 1997) revealed a bright, but unresolved
central region that produced a power-law and a MBB of 1.1 kev temperature.
The entire x-ray emitting region was found to be smaller than $\sim 460$ km. The
disk diameter at the co-rotation radius found here for this source would be
550 km. The inner disk would have to be smaller than this for a NS to produce
the high state. In a similar way, the times of ingress and egress for dips
of Cygnus X-1 (Kitamoto et al. 1984) constrain the size of the
region of origin of its low state hard spectrum to be of order 4000 km
for luminosities for which the disk inner diameter would
be near 3400 km (Misra \& Melia 1997).

Thus the inner disk-magnetopause region is a likely site of hard photon
production. This would be consistent with the large solid angles relative
to the disk seen for reflected x-rays as disk radii expand in decline.
Zhang, Yu \& Zhang (1998) have suggested that flow reversal
at the magnetopause could produce a hard power-law spectrum by the
same bulk Comptonization mechanism examined by Kluzniak \& Wilson (1991),
Hanawa (1991), and Walker (1992) for flow into a boundary layer.
Whether or not flow reversal bulk Comptonization or magnetospherically driven
return flows (Arons et al. 1984) are related to the hard spectrum,
the energy requirements for the power-law strongly suggests that it would
have to originate in the inner disk.

For both Cygnus X-1 (Balucinska-Church et al. 1997) and GRO J1655-40,
(Kuulkers et al. 1997) a soft excess was found to comprise $\sim$ 6-8\% of disk
emissions. It was modeled as a blackbody in both cases.
Temperatures of 0.13 keV and 0.60 keV, respectively, were found for the soft
excesses of the two sources. Balucinska-Church et al (1995) identified the soft
excess as disk emissions. The relatively stronger presence of the soft excess in
the deeper dips of GRO J1655-40 confirms that they must arise from the outer
disk. It is unlikely that all of the gravitational binding energy available from
the outer disk is used in producing a soft excess luminosity . The soft excess
is quite small compared to the power-law which also originates in the disk.
However if the fraction of binding energy that produces the soft excess is
roughly independent of radius, the disk temperature
will still scale as $T \sim r^{-3/4}$. The
temperature of 0.13 keV for Cygnus X-1 at 1700 km scales up to 0.58 keV
at the 230 km of GRO J1655-40. This provides additional confirmation of the
outer disk as the source of much of the soft excess.

An ultrasoft spectrum was first recognized for BHCs by White \& Marshall
(1984) for LMC X-1 and LMC X-3. Analysis of Rossi observations (Wilms et al.
1998) of these persistent, high state sources have shown that their spectra
can be fitted with a MBB (or blackbody as well for LMC X-1) of
temperature, $kT_{bb}$ = 1.0 and 1.25 keV, respectively for these sources,
along with a power law. A 5 - 9 M$_\odot$ Yilmaz NS emitting as a blackbody
would have a surface redshift of $z \sim 0.5 - 2.4$ (Fig. 1, Table 2). A local
surface temperature of 2.3 keV would produce a distantly observed apparent
temperature of T/(1+z) = 2 keV  for a low mass NS and 1.3 - 0.7 keV for the BHC.
Thus bright, ultrasoft peaks are well explained for massive NSs.

We conclude that all of the distinguishing characteristics of BHCs
are encompassed by the massive NS hypothesis. If a suitable model
for NSs were available, we might apply it with some confidence to the BHCs.
Unfortunately, none is available. Blackbody or thermal brehmsstrahlung might
be expected from near the star surfaces. Such simple models (as well as
apparently inappropriate high temperature MBBs) do a fair job
of representing the soft spectral components of both BHCs and low mass NSs.
What is lacking is an understanding of the power-law emissions. They
so dominate the low state spectra, leaving only small soft excesses, that
an optically thick, viscosity dominated disk (with its associated lower
temperature MBB spectrum) may not be an adequate starting point for
the hard x-ray spectrum. It is not certain that an optically thick
disk at the temperature of the soft excess could generate a sufficiently
energized corona to produce the power-law component.

The low state disk may be optically thin for x-rays and/or advective, i.e., not
dominated by viscosity. An advective accretion flow (ADAF) model was proposed
(Narayan et al 1997) to explain the hard quiescent radiations of BHCs. In this
model, the disk is optically thin and very hot, and accretion rates above
$10^{15-16}$ g/s are needed to produce quiescent luminosity. According
to Eq. 14, Atoll sources with magnetic fields of $\sim 2 x 10^8$ G would have
magnetospheric radii of 50 - 100 km for these accretion rates (or less,
accepting Dong Lai's suggestion). With spins of $\sim 350$ Hz they would
produce $\sim 5 x10^{34-35}$ erg/s for these accretion rates without
any surface contributions and much more
unless propeller effects block surface emissions completely. Since quiescent
luminosities of Atolls are far below these levels, the high accretion rate
ADAF model is simply inapplicable to quiescent NSs. Recall that ADAF models
with transitions to the conventional optically thick, geometrically thin,
disk (Esin, McClintock \& Narayan 1997) at radii beyond $10^4 R_g$ have
been shown ($\dot{Z}$ycki, Done \& Smith 1997) to fail for BHCs as well.
Finally, the presence of a hard spectrum for 1E1740.7-2942 for an inner
disk radius of only 225 km shows that the ADAF model simply cannot be taken
seriously. While there seems to be little to recommend pure ADAF models, a
combination of partially advective and magnetospheric effects cannot
be ruled out. An advective disk with a magnetospherically driven return
flow (Arons et al. 1984) might have some merit. Advective flows can have
positive energy for the return.

The nature of the spectrum that would arise from within the magnetosphere for
luminosities above $L_c$ is also unclear. The production of soft
photons via cyclotron emission would be difficult to quantify. The
spectrum of photons arising from the surface might be blackbody or
thermal brehmsstrahlung. Psaltis, Lamb \& Miller (1995)
have calculated spectra and successfully reproduced the color-color diagram
for a Z-source with an accretion disk-central corona-magnetosphere model.
It possible that this model might be extended to encompass the hard spectrum
of the propeller regime, for which the hot corona might be
located at the magnetopause. The hard spectrum of SAX J1808.4-3658
illustrates an additional difficulty for spectral models. The emergent
observed spectrum may be strongly affected by inclination effects.

Black hole models must necessarily use an optically thick disk
to produce soft peaks and some Comptonization of soft photons to
produce power-laws. Although current models with central black holes and
accretions disks have had some success in representing static spectra for some
states of the BHCs, most models rely on features such as a hot central corona
introduced in a purely \textit{ad hoc} fashion. It remains to be shown
rigorously that a hot central corona would develop for Keplerian flow into
an event horizon. The current "sombrero" models (Poutanen 1998)
do a good job of representing some BHC spectra but require that the
accretion disk intrusion into the central corona be ``just so''
to let just the right number of seed photons be Compton scattered to
produce the hard spectrum. It is not clear that such a model can
account for the timing signatures within the spectrum.

\section{Active Galactic Nuclei}
It must be acknowledged that active galactic nuclei pose an interesting
challenge for any alternative theory of gravity. Spectral state
switches have been observed for them (Mannheim et al 1995). They likely
do not possess star cluster cores (Maoz 1995). Massive single objects
with accretion disks might be possible but they would be most unusual
astronomical objects. It can be readily seen from Eqs. 11 and 12 that a mean
density of $\sim 1 g/cm^3$ could correspond to a very compact object of
$\sim 10^8 M_\odot$ that would not necessarily be a black hole, even in SM.
Given the extremely high efficiency of conversion of gravitational
mass-energy to luminosity, there should be no problem modeling AGN
levels of radiation. But the constant proper density model is wholly inadequate.
Realistically, an object of such low density might need to be supported
internally by radiation pressure (but see Graber 1998) and would likely have
large density gradients and strong magnetic stresses. This is clearly a topic
for another time, but one that should be pursued. Black holes would
certainly be simpler, but we should confirm them among the galactic BHCs,
where we can strongly constrain masses and radii, if at all possible.
The issue is clearly that of the existence or non-existence of an
event horizon. YM objects with $u(R) > 1/2$ are more compact than black holes
of the same mass and might well be of low luminosity with their surfaces
inside the photon orbit.

\section{Conclusions}
The characteristics of BHCs appear to be compatible with a massive NS model.
The signatures of this model are strong magnetic fields, large surface binding
energies and large redshifts of surface emissions. The field can produce
power-law and gamma emissions, jets, spectral state switches and cyclotron
emissions and the surfaces can produce ultrasoft peaks and large rms
flickering. The magnetic field strengths and spins proposed here for BHCs are
reasonably near the spin-up trend line from radio pulsars to millisecond
pulsars. Where co-rotation radii can be determined from spectral fit parameters,
the corresponding spin rates and subsequently found magnetic field strengths
imply quiescent luminosities in agreement with observations.
Due to their slower spins, the magnetospheric contributions to the
quiescent luminosities of the BHCs appear to be somehat smaller than those
of the low mass Atoll and Z NSs. The work of
$\dot{Z}$ycki, Done and Smith (1998a) describes a spectral state switch in
exact accord with the magnetic propeller mechanism cutting off the flow to the
surface with little change of inner disk radius. Cir X-1, the NS most like a
BHC, has been found to have spin and magnetic field similar to many of
the BHCs. The ubiquitous slow $\sim 6$ Hz QPOs found in high states of both BHCs
and NSs (Miyamoto, Kimura \& Kitamoto 1991, Makishima et al. 1986).
can be explained as accretion disk flow and opacity oscillations driven by
surface radiations (Fortner, Lamb \& Miller 1989). It is not known
how a black hole might produce them.

A clear implication of the massive NS model is that hard spectral features
arise from the inner disk (magnetopause?) while soft spectral features can
arise from both the disk and the surface. The strong reflected
x-rays seen in hard states with fairly large solid angles subtended
as the disk radius expands in decline, suggests that the
magnetopause might be a primary site of hard photon production.

Adoption of the massive NS hypothesis would eliminate many difficulties for
accretion disk theory. Spectral state switches would no longer require
discontinuous disk behavior for transitions between high and low states. An
additional burden for disk theory is removed by no longer requiring disk
explanations for occasional BHC failures to produce ultrasoft peaks at high
luminosities. The massive NS hypothesis explains these as either magnetic axis
orientation effects or failure of the inner disk to penetrate
the co-rotation radius. The structure of the transition region near the
magnetopause is likely to be much more complex than previously implied here.
It is likely to be involved in a complicated, inclination dependent way in
the production of the pairs of kHz QPOs.

Three of the BHC cases examined here, including Cyg X-1, seem to have
spins below 10 Hz and very strong magnetic fields.
All are above the "death line" for pulsars, while V0332+53,
the original flickering pulsar, is not. Cyg X-1 is a wind accreter
that appears to be delicately balanced between intermediate and low states.
Its large co-rotation radius permits a large energy release and consequent
large radiation pressure increase when mass does reach the star surface.
The net effect is a somewhat self-regulated luminosity.
It is likely near spin secular equilibrium, as are the Zs and Atolls.
Aside from their spins being revealed during bursts,
the Zs and Atolls are not x-ray or radio pulsars. Magnetic
fields of order $10^{10}$ G and higher may be sufficient to suppress
bursts, however, Cir X-1 has exhibited bursts.
The similarly weak magnetic fields of GRO J1655-40, A0620-00 and
GRS 1915+105 suggests that they should be carefully monitored for bursts; indeed
this last source may have already produced some (Taam, Chen, \& Swank 1997).
That the stronger-field BHCs are not bursters would be expected.
The lack of pulses requires near-coincident magnetic and spin axes or extreme
magnetic fields. Such a configuration is likely necessary for efficient
production of jets and may enhance gamma emissions.

KHz QPOs corresponding to the Keplerian frequencies of the inner disk imply
that the accretion disks of both NSs and BHCs can reach to near the
innermost marginally stable orbit. With a fixed binding energy for this
orbit it seems clear that we must look for the differences between NSs
and BHCs inside this orbit. Black hole disk models that are differentiated
from NS disks without explicitly incorporating the differences between
event horizons and star surfaces should have little credibility.

A tremendous amount of detailed modeling will be necessary to extend
the massive NS hypothesis beyond this introduction. The Yilmaz modifications
of general relativity have provided an acceptable basis for the
explorations of the differences between surfaces and event horizons. The
simplicity of the Yilmaz theory, both mathematically and conceptually, warrants
its continued use. Its strong metric similarities to general relativity
below $2 M_\odot$ and the lack of an event horizon beyond $3 M_\odot$
will provide for stringent comparisons of black hole vs NS models.
The recent discovery of NSs of approximately 2 M$_\odot$ (Zhang et al. 1998)
should permit detailed comparisons of the two theories for NSs in a range
where differences might first appear. For example, the amplitudes of QPOs
generated from bursts might be expected to differ due to differences
of gravitational curvature of photon trajectories.
Comparisons of this sort might be complicated for masses large enough for
the surfaces to be inside the photon orbit. The important point,
however, is that the Yilmaz metric provides a sharp tool that should eventually
decide the question of the existence of event horizons.

\section{Acknowledgements}
It is a pleasure to acknowledge many stimulating
and helpful discussions with my colleagues, Ray C.
Jones,  Adam Fisher and Charles Rogers. The
encouragement of Prof. Carroll O. Alley, University of
Maryland and a manuscript review by Darryl Leiter are
also appreciated. I am indebted to Adam
Fisher for technical assistance. A great debt is
owed to those whose dreams produced x-ray vision and who
have told us what they have seen.
\clearpage

\appendix{Appendix}
1. Particle Mechanics\\
\\
The mechanics of a particle orbit can be examined
very simply with the aid of the energy-momentum four-vector. The magnitude of
this vector, given by g$^{ij}$p$_i$p$_j$,
is m$_0$$^2$c$^2$ where m$_0$ is the rest mass of the
particle. For
a particle in an equatorial orbit ($\theta$ = $\pi$,
p$_{\theta}$ = 0) about
an object of gravitational mass M in the SM, one obtains:
\begin{equation}
\frac{E^{2}}{(1-2 u)c^{2}}-(1-2 u){p_{r}}^{2}-\frac{{p_{\phi}}
^{2}}{{r}^{2}}={m_{0}}^{2}c^{2}
\end{equation}\\
Here p$_{ 0}$ = E/c, where E is the particle energy and
$u = GM/c^2r$ is the gravitational potential at distance r from
the center of mass M. p$_{\phi}$, the particle angular
momentum, is a constant of the motion. Further, by
defining a = (cp$_{\phi}$/GMm$_{ 0}$) and rearranging, the equation
above becomes:
\begin{equation}
{(1 - 2 u)}^{2}\frac{{p_{r}}^{2}}{{m_{0}}^{2}c^{2}} = \frac{E^{2}}
{{m_{0}}^{2}c^{4}} - (1 - 2 u){(1 + a^{2}u^{2})}
\end{equation} For suitably small energy, bound orbits occur. Turning
points for which p$_{ r}$ = 0 can be found by examining the
effective potential, which consists of all terms to the
right of E$^{ 2}$/m$_{ 0}$$^{ 2}$c$^{ 4}$ above. At minima of the effective
potential we find
\begin{equation}
a^{2}=\frac{1}{u-3 u^{2}}
\end{equation}and with p$_{ r}$ = 0, we get
\begin{equation}
E=m_{0}c^{2}\frac{(1-2 u)}{\sqrt{{(1-3 u)}}}
\end{equation}The innermost marginally stable orbit is found by
setting the first two derivatives of the effective
potential with respect to u to zero. This gives the
well-known results, u = 1/6 and a$^{ 2}$ = 12. With this
value of u, E = $\sqrt{(8/9)}$ m$_{ 0}$c$^{ 2}$.
Substituting this value
for E, the radial motion equation for a subsequent
geodesic fall to the star surface is:
\begin{equation}
(1-2 u)\frac{{p_{r}}^{2}}{{m_{0}}^{2}c^{2}}=\frac{1}{9}(
6 u-1)^{3}
\end{equation}Equating p$_{ r}$ to mv$_{ r}$, using m = E/c$^{ 2}$
and recognizing (1 -
2u)v$_{ r}$ as the proper velocity at the position where the
potential is u, the radial velocity relative to a
distant observer is seen to be
\begin{equation}
{{v'}}_{r}=\frac{\sqrt{2}}{4}c(6 u-1)^{3/2}
\end{equation}\indent Analogous treatment of the orbit equation in the
YM yields:
\begin{equation}
\exp(-4 u)\frac{{p_{r}}^{ 2}}{{m_{0}}^{2}c^{2}}=\frac{E^{2}}{{m_{0}}^{2}c^{4}}-\exp(-2 u){(1+a^{2}u^{2}\exp(-2 u))}
\end{equation}Setting the first two derivatives of the effective
potential to zero in this case produces coupled
equations which have a solution for $u = (3 - \sqrt{5})/4$ $\approx$
0.191 with $a^2 = 12.4$. Circular orbits occur for
\begin{equation}
a^2=\frac{\exp(2 u)}{u-2 u^{2}}
\end{equation}and
\begin{equation}
E=m_{0}c^{2}\exp(-u)\sqrt{\frac{1-u}{1-2 u}}
\end{equation}A particle in geodesic infall starting with p$_{ r}$ = 0 at
the marginally stable orbit would arrive at the star
with velocity components:
\begin{equation}
{{v'}}_{r}=c{[1-1.12\exp(-2 u)(1+12.4 u^{2}\exp(-2 u)]
^{1/2}}
\end{equation}
\begin{equation}
{{v'}}_{\phi}=3.73 cu\exp(-2 u)
\end{equation}The radial velocity is supersonic for disk temperatures
of as much as 1 keV for u = 0.2; i.e., for very little
change of potential relative to that of the marginally
stable orbit. For u = 0.25, corresponding to the
surface of a star of $\approx$ 3 M$_\odot$, the
radial speed for a
Keplerian fall from the marginally stable orbit would
be about 10$^{ 9}$ cm/s.

The Keplerian frequency of the marginally stable orbit can be found
from the relations:
$ \nu_k = \frac {1} {2 \pi} \frac {d\phi}{dt} = \frac {g^{\phi \phi} p_\phi}{2 \pi m_0 c}
\frac {ds}{dt}$~,~ $p_\phi = \frac {GMm_0a}{c}$~and
$\frac{ds}{dt} = \frac {m_0 c^3}{E} \exp{(-2u)}$~ to be:
\begin{equation}
\nu_K = \frac{\sqrt{GM} exp{(-2u)}}{2 \pi r^{3/2} \sqrt{(1 - u)}}
\end{equation}

\indent In either metric there is an unstable circular
photon orbit. In the YM this lies outside
the star surface for u(R) $>$ 1/2, or for m$_{ g}$ $>$ 6 M$_\odot$. For
black holes the unstable photon orbit occurs for radius
r = 1.5 R$_{ s}$. Although radially directed photons can
always escape in the YM, those with too much
orbital angular momentum can be trapped inside the
unstable photon orbit.

2. Compact Objects\\
\\
The constant proper density object is one of
{\it constant intrinsic local density\/}. Since there are
cases of interest for extensions of this
model to variable local density,
the appropriate equations will be set up for a
more general case. Consider a spherically symmetric
density distribution $\rho_0$h(r$<$R) with
$\rho_0$ a constant and
h(r) a distribution function for local density. The
condition h(0) = 1 is imposed without loss of
generality, and all mass is confined within radius R
relative to a distant observer. In this model, the
proper density, $\rho$(r)g(r) is assumed to be given by
$\rho_0$h(r). Thus it will be assumed that
$\rho(r) = \rho_0h(r)/g(r)$. As shown by Clapp (Clapp
1973), u(r) can be constructed by adding the potentials of successive
shells of matter. Inside each shell the contribution
to u(r) is constant and outside it declines as 1/r. 
For a general radius r, with r $<$ R, contributions to
the potential must be separated into those arising from
shells of radius $r'$ inside r and those from shells of
radius $r'$ outside r. The gravitational mass within
each shell is $4 \pi r'^2 \rho_0 h(r')dr'/g(r')$. (For the free
baryon mass substitute $g(r')^{3/2}$.) The central potential
u(0) is obtained by dividing each shell mass increment
by $r'$ and integrating out to R. Subtracting u(0) from
u(r$<$R) leads to Clapp's integral equation:
\begin{equation}
u(r<R)-u(0)=\frac{-4\pi G \rho_0}{c^2}\int_{0}^{r}{{dr'}
h({r'}){({r'}-{{r'}}^{2}/r)}/g({r'})}
\end{equation}
It can be shown that this equation is a solution of the
field equations of either gravitational theory by
substitution of the appropriate g(r) and other metric
components into the field equations and solving the G$_{tt}$
differential equation for u(r). By expressing u(r$<$R)
and u(0) in terms of g(r) and g(0) and incorporating
g(0) into a scale factor for lengths, the integral can
be recast in the dimensionless form:
\begin{equation}
y(x)=-2\int_{0}^{x}{f({x'})h({x'})({x'}-{{x'}}^{2}/x){dx'}}
\end{equation}
where x = r/(r$_{\rm 0}$g(0)$^{\rm N/2}$),
r$_0$ = $c/\sqrt{4 \pi G \rho_0}$, f(r) = g(0)/g(r), N = 1
in the YM and N = 2 in the SM
Thus these collapsed objects can
be described with the aid of the characteristic radius,
r$_0$, and a characteristic mass, M$_0$ = c$^2r_0/G$.
In the exponential metric:
\begin{equation}
y(x)=\ln(f(x))
\end{equation}and in the SM:
\begin{equation}
y(x)=1-1/f(x)
\end{equation}

After substituting the appropriate left member, the integral equation
can be numerically solved for f(x) by an iterative
relaxation method starting from f(x) = 1 for x $<$ 1 and
an asymptotic value for x $>$ 1. Asymptotic solutions
for the constant intrinsic density model, for which
h(r) = 1 are f(x) = $x^{-2}$ for the YM, (Clapp 1973)
and $x^{-1}$ found here for the SM. Physical
quantities such as gravitational mass $m_g$, equivalent
free baryon mass,$m_b$, distantly observed red shift of
surface radiations, z and radius, R can be expressed in
terms of certain integrals over f(x). Following Clapp
(Clapp 1973), these are:
\begin{equation}
F_{1}(x)=\int_{0}^{x}{h({x'})f({x'}){x'}{dx'}}
\end{equation}
\begin{equation}
F_{2}(x)=\frac{1}{x}\int_{0}^{x}{h({x'})f({x'}){{x'}}^{2}{
dx'}}
\end{equation}
\begin{equation}
F_{3}(x)={\frac{1}{x}}\int_{0}^{x}{f({x'})^{3/2}{{x'}}^{2}
h({x'}){dx'}}
\end{equation}In the exponential metric there follows:
\begin{equation}
m_{g}=M_{0}xF_{2}(x)\exp(-F_{1}(x))
\end{equation}
\begin{equation}
m_{b}=M_{0}xF_{3}(x)
\end{equation}
\begin{equation}
R=r_{0}x\exp(-F_{1}(x))
\end{equation}
\begin{equation}
u(0)=F_{1}(x)
\end{equation}
\begin{equation}
u(R)=F_{2}(x)
\end{equation}In the SM the same quantities are
given by:
\begin{equation}
m_{g}=M_{0}\frac{xF_{2}(x)}{(1+2 F_{1}(x))^{2}}
\end{equation}
\begin{equation}
m_{b}=M_{0}\frac{xF_{3}(x)}{(1+2 F_{1}(x))^{3/2}}
\end{equation}
\begin{equation}
R=r_{0}\frac{x}{(1+2 F_{1}(x))}
\end{equation}
\begin{equation}
u(0)=\frac{F_{1}(x)}{1+2 F_{1}(x)}
\end{equation}
\begin{equation}
u(R)=\frac{F_{2}(x)}{1+2 F_{1}(x)}
\end{equation}

Numerical solutions of Eq. 14 and the appropriate
integrals for the case h(r) = 1 were used to generate
the data of Table 2 and Table 3. Plots of mass, radius
and redshift for surface emissions are shown in Figures
1 and 2. Choosing h(r) = 1 forces the object to be of
{\it constant local intrinsic density\/}. To an external
observer, there would be substantial variations of
density, but locally everything would seem entirely
normal. Such is life in radically curved space-time. 
Simple extensions of this approach to an Oppenheimer -
Volkoff method for the determination of $\rho_0$
and h(r) are
possible, as well as exponential atmosphere models of
active galactic nuclei (h(r) not equal to 1), but the essential differences
between black holes, neutron stars and nuclear-density
compact objects in the exponential metric are clearly
revealed by this simple model.

\clearpage

\makeatletter
\def\jnl@aj{AJ}
\ifx\revtex@jnl\jnl@aj\let\tablebreak=\nl\fi
\makeatother

\begin{deluxetable}{lllllllll}
\tiny
\tablecaption{Neutron Star Properties}
\tablehead{
\colhead{Object} &  \colhead{$L_{min,36}$} & \colhead{$L_{c,36}$} &
  \colhead{$L_{q,33}$} & \colhead{M ($M_\odot$)} & \colhead{R (km)\tablenotemark{a}} &
  \colhead{$\nu_S$ (Hz)} & \colhead{$B_9$(G)} &
  \colhead{References.} }

\startdata
Neutron Stars (Pulsars) & & & & & & & & \nl
V0332+53 & 0.07 & & & 1.4 & 15 & .227 & 270 & 12 \nl
PSR GX 1+4 & & & & 1.4 & 15 & 0.0083 & 31000 & 13 \nl
PSR GRO J1744-28 & 1.8 & & & 1.4 & 15 & 2.14 & 103 & tw, 13 \nl
PSR 1055-52 & & & 0.2 & 1.4 & 15 & 5 & 1500 & tw, 32 \nl
Sax J1808.4-3658 & 1.5 & 0.4 & 0.17 \tablenotemark{a} & 1.4 & 15 & 401\tablenotemark{b} & 0.21 \tablenotemark{a} & 1, 2 \nl
Neutron Stars (Atoll)  & & & & & & & & \nl
Aql X-1 & 1.2 & & 0.4 & 1.4 & 15 & 549\tablenotemark{b} & 0.13\tablenotemark{a} & 3, 4, 20 \nl
4U 1608-52 & 10 & & 2 & 1.4 & 15 & 460\tablenotemark{b} & 0.46\tablenotemark{a} & 7, 20, tw \nl
4U 1820-30 & 15 & & 1-3.4\tablenotemark{a} & 1.9 & 15 & 275-550\tablenotemark{b} & 0.5-1.1 & 5, 6 \nl
4U 1730-335 & 10 & & 2 & 1.4 & 15 & 566\tablenotemark{a} & 0.36\tablenotemark{a} & 7, tw \nl
4U 2129+47 & 6.3 & & 0.63 & 1.4 & 15 & 374\tablenotemark{a} & 0.47\tablenotemark{a} & 7, 20, tw \nl
KS 1731-26 & 10 & & 1.8\tablenotemark{a} & 1.4 & 15 & 524\tablenotemark{b} & 0.4\tablenotemark{a} & 7, tw \nl
Neutron Stars (Z) & & & & & & &  & \nl
Cyg X-2 & 100 & & 10\tablenotemark{a} & 1.9 & 15 & 346 & 2.2\tablenotemark{a} & 8, tw \nl
Sco X-1 & & & 7.2\tablenotemark{a} & 1.4 & 15 & 310 & 2.3 & tw, 9 \nl
GX 5-1 & & & 8.5\tablenotemark{a} & 1.4 & 15 & 330 & 2.2 & tw, 9 \nl
GX 17+2 & & & 6.9\tablenotemark{a}& 1.4 & 15 & 306 & 2.3 & tw, 9 \nl
Cir X-1 & 630 & 30 & 0.9\tablenotemark{a} & 1.4 & 15 & 32\tablenotemark{a} & 80\tablenotemark{a} & tw, 10,11 \nl
Cir X-1 & 630 & 30 & 6\tablenotemark{a} & 7 & 18 & 38\tablenotemark{a} & 82\tablenotemark{a} & tw, 6 \nl
  & & & & & & & & \nl
Black Hole Candidates  & & & & & & & & \nl
GRS 1124-68 &  & 6.6 & 0.39 \tablenotemark{a} & 5 & 18.7 & 15\tablenotemark{a} & 112\tablenotemark{a} & tw, 14, 15, 16 \nl
GS 2023+338 & 1008 & 48\tablenotemark{a} & 9 & 7 & 18 & 39\tablenotemark{a} & 95\tablenotemark{a} & tw, 18, 19, 20 \nl
1E1740.7-2942 & & 30 & $<7$\tablenotemark{a} & 7? & 18 & $<43$\tablenotemark{a} & $>69$\tablenotemark{a} & tw, 21 \nl
A0620-00 & 8 &  & .01 & 4.9 & 18.7 & 13.5\tablenotemark{a} & 23\tablenotemark{a} & tw, 20, 24 \nl
GRO J1655-40 & 31 & & 0.25 & 7 & 18 & 34\tablenotemark{a} & 21\tablenotemark{a} & tw, 20, 25, 26, 27 \nl
GRS 1915+105 & 1600 & 19\tablenotemark{a} & 7\tablenotemark{a} & 7\tablenotemark{a} & 18 & 67\tablenotemark{b} & 28\tablenotemark{a} & tw, 29, 30, 31 \nl
GS 2000+25 &  & 0.15 & $0.001-0.004$\tablenotemark{a} & 7 & 18 & 4.5-1.6\tablenotemark{a} & 140-650\tablenotemark{a} & tw, 17, 18 \nl
Cygnus X-1 & & 50 & 1\tablenotemark{a} & 10 & 14 & 2.6\tablenotemark{a} & 15000\tablenotemark{a} & tw, 22, 23 \nl
GRO J0422+32 &?1.5 & & $<0.079$ & 3.6 & 18 & ?0.23 \tablenotemark{a} & ?1000 \tablenotemark{a} & tw, 20, 28 \nl
\tablenotetext{a}{Values calculated in this work.}
\tablenotetext{b}{QPO frequency.}
\tablerefs{tw = this work, 1-Gilfanov et al. 1998,
2-Heindl \& Smith 1998, 3-Campana et al 1998b, 4-Zhang, Yu \& Zhang 1998,
5-Zhang et al 1998, 6-Swank 1998, 7-Campana et al 1998a,
8-Casares, Charles \& Kuulkers 1998, 9-White \& Zhang 1997,
10-Dower, Bradt \& Morgan 1982, 11-Bradt, Shirey \& Levine 1998,
12-Stella, White \& Rosner 1986, 13-Cui 1997, 14-Misra \& Melia 1997,
15-Ebisawa et al. 1994, 16-$\dot{Z}$ycki, Done \& Smith 1998,
17-$\dot{Z}$ycki, Done \& Smith 1997, 18- Tanaka \& Lewin 1995,
19-Tanaka 1992, 20-Garcia, McClintock \& Narayan 1997
21-Vilhu et al. 1997, 22-Belloni et al. 1996, 23-Misra \& Melia 1997,
24-Kuulkers 1998, 25-Bailyn et al 1995, 26-Remillard et al 1997,
27-Mendez et al 1997, 28-Grove et al. 1998, 29-Belloni et al 1997,
30-Remillard et al 1997, 31-Remillard \& Morgan 1998,
32-Shibata et al. 1997
 }

\enddata
\normalsize
\end{deluxetable}

\makeatletter
\def\jnl@aj{AJ}
\ifx\revtex@jnl\jnl@aj\let\tablebreak=\nl\fi
\makeatother

\begin{deluxetable}{lllllllll}
\tablecaption{Quantities in the Yilmaz Metric}
\tablehead{
\colhead{x   } &  \colhead{f(x)} &   \colhead{m$_b$} &
           \colhead{m$_g$} &  \colhead{R} &      \colhead{z} &
           \colhead{u(R)} &   \colhead{u(0)} &   \colhead{gr$_0/c^2$}}

\startdata
~0.150 &  0.993 &  0.001 &  0.001 &  0.148 &  0.009 &  0.009 &  0.013 &  0.061 \nl
~0.250 &  0.980 &  0.006 &  0.006 &  0.242 &  0.023 &  0.023 &  0.033 &  0.098 \nl
~0.350 &  0.961 &  0.015 &  0.014 &  0.328 &  0.044 &  0.043 &  0.063 &  0.138 \nl
~0.450 &  0.936 &  0.030 &  0.028 &  0.406 &  0.071 &  0.069 &  0.102 &  0.182 \nl
~0.550 &  0.907 &  0.053 &  0.047 &  0.474 &  0.105 &  0.100 &  0.149 &  0.234 \nl
~0.650 &  0.873 &  0.085 &  0.072 &  0.531 &  0.145 &  0.136 &  0.203 &  0.293 \nl
~\tablenotemark{a} 0.657 &  0.871 &  0.087 &  0.074 &  0.533 &  0.148 &  0.138 &  0.207 &  0.297 \nl
~0.750 &  0.837 &  0.124 &  0.101 &  0.576 &  0.191 &  0.175 &  0.264 &  0.361 \nl
~0.950 &  0.758 &  0.229 &  0.167 &  0.637 &  0.299 &  0.262 &  0.400 &  0.534 \nl
~1.25  &  0.635 &  0.441 &  0.269 &  0.666 &  0.497 &  0.403 &  0.630 &  0.906 \nl
~1.55  &  0.520 &  0.697 &  0.353 &  0.649 &  0.724 &  0.545 &  0.871 &  1.45 \nl
~1.85  &  0.419 &  0.974 &  0.413 &  0.611 &  0.966 &  0.676 &  1.11  &  2.18 \nl
~2.05  &  0.362 &  1.16  &  0.439 &  0.581 &  1.13  &  0.755 &  1.26  &  2.76 \nl
~2.85  &  0.203 &  1.83  &  0.475 &  0.477 &  1.71  &  0.996 &  1.79  &  5.65 \nl
~3.55  &  0.127 &  2.29  &  0.466 &  0.415 &  2.07  &  1.12  &  2.15  &  8.29 \nl
~4.25  &  0.081 &  2.59  &  0.454 &  0.384 &  2.26  &  1.18  &  2.41  &  10.1 \nl
~7.05  &  0.023 &  3.33  &  0.406 &  0.332 &  2.39  &  1.22  &  3.05  &  12.5 \nl
~10.1  &  0.010 &  3.70  &  0.383 &  0.328 &  2.22  &  1.17  &  3.42  &  11.5 \nl
~15.1  &  0.004 &  4.05  &  0.368 &  0.340 &  1.95  &  1.08  &  3.79  &  9.40 \nl
~20.0  &  0.002 &  4.27  &  0.364 &  0.355 &  1.78  &  1.02  &  4.03  &  8.00 \nl
~40.1  &  0.001 &  4.80  &  0.372 &  0.401 &  1.53  &  0.928 &  4.62  &  5.85 \nl
~60.1  &  0.000 &  5.13  &  0.384 &  0.422 &  1.49  &  0.912 &  4.96  &  5.38 \nl
~80.1  &  0.000 &  5.39  &  0.394 &  0.430 &  1.50  &  0.916 &  5.23  &  5.31 \nl
~100   &  0.000 &  5.60  &  0.401 &  0.434 &  1.52  &  0.924 &  5.44  &  5.39 \nl
~200   &  0.000 &  6.29  &  0.418 &  0.434 &  1.62  &  0.963 &  6.13  &  5.81 \nl
~400   &  0.000 &  6.98  &  0.425 &  0.434 &  1.67  &  0.982 &  6.83  &  6.04 \nl
~1000. &  0.000 &  7.90  &  0.431 &  0.434 &  1.70  &  0.993 &  7.74  &  6.17 \nl
\tablenotetext{a}{Entries for 1.4 M$_\odot$ with $\rho_0 = 1.35x10^{14} g/cm^3$.}
\tablecomments{For x, f(x), see appendix. R is in units of r$_0 =
c/\sqrt{4 \pi G \rho_0}$, m$_b$ and m$_g$ in units
of $M_0 = c^{2} r_0 /G$. g is proper surface free-fall acceleration.}
\enddata
\end{deluxetable}

\makeatletter
\def\jnl@aj{AJ}
\ifx\revtex@jnl\jnl@aj\let\tablebreak=\nl\fi
\makeatother

\begin{deluxetable}{lllllllll}
\tablecaption{Quantities in the Schwarzschild Metric}
\tablehead{
\colhead{x   } &  \colhead{f(x)} &   \colhead{m$_b$} &
           \colhead{m$_g$} &  \colhead{R} &      \colhead{z} &
           \colhead{u(R)} &   \colhead{u(0)} &   \colhead{gr$_0/c^2$}}

\startdata
~0.143 &  0.993 &  0.001 &  0.001 &  0.140 &  0.007 &  0.007 &  0.010 &  0.050 \nl
~0.288 &  0.973 &  0.007 &  0.007 &  0.266 &  0.027 &  0.026 &  0.038 &  0.099 \nl
~0.498 &  0.925 &  0.028 &  0.026 &  0.401 &  0.071 &  0.062 &  0.097 &  0.172 \nl
~0.653 &  0.880 &  0.050 &  0.044 &  0.466 &  0.111 &  0.095 &  0.143 &  0.226 \nl
~0.803 &  0.832 &  0.073 &  0.061 &  0.505 &  0.149 &  0.122 &  0.185 &  0.277 \nl
~0.918 &  0.793 &  0.090 &  0.073 &  0.524 &  0.178 &  0.140 &  0.214 &  0.314 \nl
~\tablenotemark{a} 0.923 &  0.792 &  0.091 &  0.074 &  0.525 &  0.180 &  0.141 &  0.215 &  0.316 \nl
~1.003 &  0.765 &  0.102 &  0.081 &  0.534 &  0.199 &  0.152 &  0.234 &  0.341 \nl
~1.153 &  0.716 &  0.122 &  0.093 &  0.545 &  0.231 &  0.170 &  0.264 &  0.385 \nl
~1.353 &  0.655 &  0.143 &  0.104 &  0.550 &  0.269 &  0.190 &  0.297 &  0.438 \nl
~1.513 &  0.609 &  0.157 &  0.111 &  0.550 &  0.295 &  0.202 &  0.318 &  0.475 \nl
~1.653 &  0.573 &  0.168 &  0.115 &  0.549 &  0.314 &  0.210 &  0.334 &  0.503 \nl
~1.853 &  0.526 &  0.180 &  0.120 &  0.546 &  0.336 &  0.220 &  0.353 &  0.539 \nl
~2.053 &  0.484 &  0.190 &  0.123 &  0.542 &  0.354 &  0.227 &  0.368 &  0.569 \nl
~2.353 &  0.430 &  0.201 &  0.126 &  0.536 &  0.375 &  0.236 &  0.386 &  0.604 \nl
~2.753 &  0.373 &  0.211 &  0.128 &  0.529 &  0.393 &  0.242 &  0.404 &  0.639 \nl
~3.053 &  0.338 &  0.216 &  0.129 &  0.524 &  0.403 &  0.246 &  0.414 &  0.658 \nl
~3.553 &  0.291 &  0.222 &  0.129 &  0.518 &  0.413 &  0.250 &  0.427 &  0.680 \nl
~4.053 &  0.255 &  0.226 &  0.129 &  0.514 &  0.419 &  0.252 &  0.437 &  0.695 \nl
~4.998 &  0.206 &  0.231 &  0.129 &  0.508 &  0.424 &  0.253 &  0.449 &  0.710 \nl
~6.050 &  0.168 &  0.233 &  0.129 &  0.504 &  0.429 &  0.255 &  0.458 &  0.725 \nl
~8.050 &  0.125 &  0.235 &  0.128 &  0.501 &  0.427 &  0.254 &  0.469 &  0.725 \nl
~10.05 &  0.100 &  0.235 &  0.127 &  0.500 &  0.425 &  0.254 &  0.475 &  0.722 \nl
~15.05 &  0.066 &  0.236 &  0.126 &  0.500 &  0.421 &  0.252 &  0.483 &  0.717 \nl
~20.00 &  0.050 &  0.236 &  0.126 &  0.500 &  0.419 &  0.252 &  0.488 &  0.713 \nl
\tablenotetext{a}{Entries for 1.4 M$_\odot$ with $\rho_0 = 1.35x10^{14} g/cm^3$.}
\tablecomments{For x, f(x), see appendix. R is in units of r$_0 =
c/\sqrt{4 \pi G \rho_0}$, m$_b$ and m$_g$ in units
of $M_0 = c^{2} r_0 /G$. g is proper surface free-fall acceleration.}
\enddata
\end{deluxetable}

\end{document}